\input mn
\input psfig.tex
\hsize=126mm

\vsize=195mm
\def\frac#1#2{{\begingroup#1\endgroup\over#2}}

\begintopmatter
\title{Flattened Galactic Haloes and Baryonic Dark Matter}	

\author{Srdjan Samurovi\'c,$^{1}$
Milan M.~\'{C}irkovi\'{c}$^{2,3}$ and
Vesna Milo\v sevi\'c-Zdjelar$^4$
}      
	

\affiliation{$^1$ Dipartimento di Astronomia, Universit\`{a} degli Studi di Trieste, 
 Via Tiepolo 11, I-34131 Trieste,
ITALY \break 
srdjan@ts.astro.it	
}

\smallskip
\affiliation{$^2$ Astronomical Observatory, Volgina 7, 11000 Belgrade, SERBIA}

\smallskip
\affiliation{$^{3}$ Dept. of Physics \& Astronomy, 
SUNY at Stony Brook, 
Stony Brook, NY 11794-3800, USA\break
 cirkovic@sbast3.ess.sunysb.edu}
\smallskip
\affiliation{$^{4}$ Dept. of Physics, University of Manitoba,
Winnipeg MB,  R3T 2N2, CANADA} 

\shortauthor{S. Samurovi\'c, M. M. \'Cirkovi\'c and V. Milo\v sevi\'c-Zdjelar}		
\shorttitle{Flattened Galactic Haloes and Baryonic Dark Matter}
	
\bigskip 

\abstract{We discuss the tight interconnection between microlensing optical depths, 
flattening of dark haloes and low-to-intermediate redshift baryonic census. 
By analysing  plots of the microlensing optical depth 
as a function of galactic coordinates  for different values 
of axis ratio $q$ of the galactic MACHO halo, we have shown  that observations
are best described by a flattened halo  with  $q \simeq 0.6$. There is no dynamical obstacle for such a choice of global halo shape.
Both extremely flattened $q \simeq 0.2$ and spherical $q \simeq 1$ haloes have several difficulties, although not of equal severity. 
Consequences of such flattening for the cosmological density fraction contained
in MACHOs  are considered and comparison with mass in low and
intermediate-redshift Ly$\alpha$ forest and other plausible reservoirs of 
gas is discussed in context of a unified description of the evolution of 
baryonic content of the universe. }


\maketitle

\section{Introduction}
The question of the nature and properties of the baryonic dark matter  
(BDM) is one of the most active fields of recent astrophysical research
(Hegyi \& Olive 1986; 
Persic \& Salucci 1992, 1998; Ashman 1992; Gnedin \& Ostriker 1992; Richstone et al. 1992; 
Carr 1994 and references therein; Wasserman \& Salpeter 1994; 
Flynn, Gould \& Bahcall 1996; 
Graff \& Freese 1996), because of its impact on the various branches of the modern astrophysics and cosmology. Microlensing searches 
have proved to be one of the most important
tools for investigation of properties of the halo of the Milky Way 
(Paczy\'nski 1986; Griest et al. 1991; De R\'ujula, Jetzer \& Mass\'o 1992; 
Paczy\'nski et al. 1994; Aubourg et al. 1993; 
Sackett \& Gould 1993; Gould 1994, 1996; Alcock et al. 1996, 1997a, b, c; 
Ansari et al. 1996).  
Comparison of theoretical models and microlensing (ML) data has
already yielded intriguing results and insights (Gates, Gyuk \& Turner
1995; Steigman \& Tkachev 1998). Under
the Copernican assumption that the Milky Way is a typical zero-redshift $L_\ast$
galaxy, it is natural to ask what consequences recently discovered
Massive Compact Halo Objects (MACHOs) have for the global picture of
evolution of the baryonic structure in the universe. 

By MACHOs we denote present-day collapsed objects residing in the halo of the Milky 
Way (and, by a Copernican assumption, haloes of other normal spirals)
that are baryonic in nature and/or origin.
Thus, we exclude hypothetical primordial mini black holes (e.g. Canuto 1978)
and exotic
non-baryonic aggregates (e.g. Kolb \& Tkachev 1995;
 Eichler 1996). The motivation behind this is the crucial importance of 
the Big Bang nucleosynthesis (BBNS) constraints for cosmological density fraction
$\Omega_B \equiv \rho_B / \rho_{\rm crit}$ in baryons (Yang et al. 1984; Walker et al. 
1991; but see also Gnedin \& Ostriker 1992; Hata et al. 1996).  Objects detectable
only through microlensing searches may present an important, and indeed a dominant,
item in the total baryonic census.

We hereby intend to extend the discussion of an important recent paper
by Fields, Freese \& Graff (1998; hereafter FFG). While they approach
the problem of cosmological density fraction $\Omega_{\rm MACHO}$
contained in collapsed halo objects from the point of view of
mass-to-light ratio and MACHOs as products of stellar evolution,
we take a slightly different approach to achieve similar ends. In our view,
microlensing observations give the most significant key to the 
spatial distribution of baryonic matter in haloes of 
large spiral galaxies like the Milky Way. Under the assumption
that our galaxy is typical for the normal galactic population at
zero redshift, we apply integration of spherical and   various flattened  model
MACHO distributions over the luminosity function in order to 
determine cosmological contribution of these dark haloes. We also
extend the discussion of FFG on the comparison between MACHO and
Ly$\alpha$ forest mass density, especially in light of recent
observational indications that most of the low-redshift Ly$\alpha$ 
forest is associated with galaxies (Spinrad et al. 1993; 
Lanzetta et al. 1995; Chen et al. 1998; Yahata et al. 1998). 
This conjecture is crucial 
to the understanding of transition between different types of baryonic 
dark matter, which does not seem to occur later than the differentiation of
galactic structure itself. This is in accordance with the
cooling flow-type models of galaxy formation (Nulsen \& Fabian 1995, 
1997), and offer simple interpretation of the best
available cosmic baryon census (Fukugita, Hogan \& Peebles 1998,
hereafter FHP), strengthening the assumption of FFG that MACHOs were
formed before the Ly$\alpha$ systems. Thus, our paper is 
complementary to FFG in several ways. 

Global flattening of the galactic dark halo  has been recently considered in
detail by a number of authors (Binney, May \& Ostriker 1987; Griest et al. 1991; 
Sackett \& Gould 1993; Frieman \& Scoccimarro 
1994; Nakamura, Kan-ya \& Nishi 1996). 
We use and update their results, especially
excellent discussions of Sackett \& Gould (1993) and 
Frieman \& Scoccimarro (1994) which, unfortunately, were written before the bulk
of the existing microlensing data came, in order to emphasize the tight relation
between the global shape of the baryonic halo and its total mass, with all
its cosmological implications. The goal of this paper, accordingly, is (1) to update
the discussion of discrimination of various halo models according to microlensing
optical depths produced, and (2) to investigate the influence of realistic global flattening, 
 constrained in such a way,
on estimates of baryonic
cosmological density and consequences for the evolution of the baryonic
content of the universe. Thus, inclusion of more phenomena, like the global flattening, 
present another step towards a unified picture of the physical processes of relevance to 
the baryonic matter.

\section{Baryonic dark matter in  the Galaxy: a short overview}

{}From shape of the Milky Way rotation curve (RC) 
(Merrifield 1992) one can see that a huge 
amount of mass still has to be identified.  The difficulties 
in the determination of the RC led to  uncertainties in the most 
important parameters such as the galactic constant $R_0$, which represents the 
distance to the Galactic centre, and  the circular speed at the Solar 
radius, $v_0$ (Merrifield 1992; Olling \& Merrifield 1998; Sackett 1997).
Although the IAU 1986 standard values are $R_0=8.5$ kpc and $v_0=220\, {\rm km s^{-
1}}$ some recent estimates allow the smaller values: $R_0=7.1\pm 0.4$ kpc and 
$v_0=184\pm 8\, {\rm km s^{-1}}$ (Olling \& Merrifield 1998). In this paper 
we adopt the value $R_0=8.5\pm 
0.5$ kpc (Feast \& Whitelock 1997) based upon an analysis of {\it HIPPARCOS} proper 
motion of 220 Galactic Cepheids and $v_0=210\pm 25\, {\rm km s^{-1}}$  that 
includes the best values from the H{\sixrm I} analysis ($v_0=185\, {\rm km s^{-1}}$)
and the estimated value based on the Sgr A* proper motion $v_0=235\, 
{\rm km s^{-1}}$ (Sackett 1997). 

Without going into the  discussions about the content of the dark matter in the halo, we
only state here that one part (presumably smaller) has to be in the baryonic 
form. Namely, cosmic nucleosynthesis predicts that (Turner 1996; FHP): 
$$
0.0062\;\le  \;\Omega_B\; h^2\; \le \;0.026\eqno (1)
$$ 
where $\Omega_B$ is the universal baryonic mass-density parameter 
($\Omega_B\equiv {\rho_B / \rho_{\rm crit}}=8\pi 
G\rho _B/3H_0^2$) and  $ 0.4^< _\sim\; {h}\;^<_\sim 1.0$. The "silent" $h$ is 
used in parametrization of the Hubble constant $H_0=100 \, h\; {\rm km\,s^{-
1}\,Mpc^{-1}}$. Recent measurements of the primordial deuterium abundance
 (Burles \& Tytler 1998) give: 
 $$\Omega_B h^2=0.0193. \eqno (2)$$ 

Using the simplest dynamical estimate of the mass of the Galaxy (Kepler's 
third  law) one can obtain
 (Roulet \& Mollerach 1997, and references therein):
$$\eqalign{M_{\rm dyn} & \simeq  {v_c^2r_{\rm max}\over G} \simeq \cr
& \simeq  5.6\times 10^{11}
\left ({v_c\over 220 \; {\rm km/s}}\right )^2 \! \left ( 
{r_{\rm max}\over 50 \; {\rm kpc}}\right ){\rm M_\odot} ,}\eqno(3)$$ 
where $M(r)$ is the mass interior to 
$r_{\rm max}$, $v_c$ is the measured rotational velocity and $r$ is the radius within 
which most of the dynamical mass  of the  Galaxy is located.  For the entire visible 
matter (stars, interstellar and intracluster gas) one
 can obtain cosmological density fraction (Persic \& Salucci 1992; see also Bristow \& Phillipps 1994): 
$$
\Omega_{\rm vis} \approx 2.2 \times 10^{-3} + 6.1\times 10^{-4} \, h^{-1.3}. \eqno(4)
$$
The discrepancy between the Eqs.~(1) and (4) represents the so-called problem of missing baryons.
Various types of such dark baryonic  material have been suggested: 
gaseous clouds of plasma or neutral atoms and molecules, snowballs or icy 
bodies similar to comets, stars, planets, white dwarfs, neutron stars and 
stellar or primordial black holes (e.g. Hegyi \& Olive 1986; Peebles 1993). \bigskip

On the other hand, the dynamical mass in the Eq.~(3) corresponds to the cosmological density
parameter (if the Milky Way is a typical $L_\star$ galaxy).
$$
\Omega_{\rm dyn} \simeq 0.063h \!
\left ({v_c\over 220 \; {\rm km/s}}\right )^2 \! \left ( 
{r_{\rm max}\over 50 \; {\rm kpc}}\right ),\eqno(3.1)$$ 
(This result depends on details of the Galaxy luminosity function which will be discussed in the Section 5).
The discrepancy between the Eqs.~(3.1) and (4) can be regarded as a formulation of the general dark matter
problem on the galactic scales.
The necessity for the dark matter is emphasized by 
severe limits on mass-to-light ratio in the Local
Group imposed by deep blank sky surveys (Richstone et al. 1992; Hu
et al. 1994; Flynn, Gould \& Bahcall 1996), as well as
with huge dynamical mass for the Milky Way inferred by Kulessa \&
Lynden-Bell (1992) and Lee et al. (1993).

 The mass in the halo is dominated by the matter that is not, at least easily, 
detectable. 
According to the recent observations of satellite galaxies (Zaritsky et al. 1997)
dark haloes extend to much larger radius than 50 kpc.
So, one can formally write: 
$$\Omega_{\rm dyn}\sim 0.1 \; {}^>_\sim 15\, \Omega_{\rm vis}.\eqno(5)$$

\section{Microlensing and flattening}

In searches for the BDM content the method of microlensing has so far 
proved successful. Its name derives from the fact that  lensing of distant 
objects is made by bodies with masses characteristic of a star or planet. 
Although the theoretical development of this idea started in 1964 (e.g. 
Peebles 1993, and references therein), it was the seminal paper by Paczy\'nski  
(1986) that showed that one can search for ML events in the Milky Way halo if 
it is made of stars or brown dwarfs. Rapid development of observational and 
computer technology led to the detection of  a significant number of ML events 
(e.g. Mellier, Bernardeau \& Van Waerbeke 1998).  Searches have been directed towards
Large  and Small Magellanic Clouds (LMC and SMC) (Alcock et al.~1996, 1997b; Ansari et al.
1996; Palanque-Delabrouille et al. 1998), Galactic bulge (Kiraga \& Paczy\'nski 
1994) and M31 (Crotts 1992, Crotts \& Tomaney 1996; see also Gould 1994).

All these surveys give results concerning two important parameters: masses 
of the intervening lenses and the number of  lenses within the Einstein radius around the
line of sight to a lensed source---the  optical depth. In Table 1 we give the 
targets observed, names of the appropriate survey, mass ranges of the 
lenses, and corresponding optical depth.  

\begintable*{1}
\caption{{\bf Table 1.}  Targets in different ML surveys, the mass 
ranges of the lenses and optical depths. References: 
(1) Alcock et al. 1997b, (2) Renault et al. 1997, 
(3) Alcock et al. 1997a,
(4) Crotts \& Tomaney 1996,
(5) Palanque-Delabrouille et al. 1998. }
\halign{%
\rm#\hfil&\qquad\rm#\hfil&\qquad\rm\hfil#&\qquad\rm\hfil
#&\qquad\rm\hfil#&\qquad\rm\hfil#&\qquad\rm#\hfil
&\qquad\rm\hfil#&\qquad\rm#\hfil&\qquad\hfil\rm#\cr
Target & Survey & Mass range & Optical depth\cr
 {\rm LMC/SMC}$^1$  & {\sevenrm MACHO} & $\approx$ 0.3 -- 0.5 ${\rm M_\odot}$
  & $\tau_{\rm LMC}=2.9^{+1.4}_{-0.9}\times 10^{-7}$\cr
 & & &$\tau_{\rm SMC}=1.5-3\times 10^{-7}$\cr
 {\rm LMC}$^2$  & {\sevenrm EROS} & 0.1 ${\rm M_\odot}$ & $\tau_{\rm LMC}=0.82_{-0.5}^{+1.1}\times
10^{-7}$\cr
Gal. bulge$^3$ & {\sevenrm MACHO;DUO;OGLE}&0.08 -- 0.6 ${\rm M_\odot}$ & 
$\tau_{\rm bul}=3.9^{+1.8}_{-1.2}\times 10^{-6}$&\cr  
 {\rm M31}$^4$& {\sevenrm KPNO} & $\approx$ 1.0 ${\rm M_\odot}$ & \ \ $\tau_{\rm M31}
=5-10\times 10^{-6}$ &\cr
 {\rm LMC/SMC}$^5$& {\sevenrm EROS2} & $2.6_{-2.3}^{+8.2}$ ${\rm M_\odot}$ &
 $\tau _{\rm SMC}=3.3\times 10^{-7}$&\cr
}
\endtable
 It is known from 
the work of Sackett \& Gould (1993) that  instead  of the equation 
for the mass density in a   spherical halo: 
$$
\rho ({\bf r})={v_\infty^2 \over 4\pi G}\left ({1\over r_c^2+r ^2}\right 
)\theta (R_T-r)\eqno (6)
$$
(where $r$ is the Galactocentric radius, $v_\infty$ 
is the asymptotic circular speed of the halo, $r_c$ is the core radius of the 
halo and $R_T$ is the truncation radius), one should use the general formula
for flattened halo: 
$$
\rho ({\bf 
r})={\tan \psi\over \psi}{v_\infty^2 \over 4\pi G}\left ({1\over r_c^2+\zeta 
^2}\right )\theta (R_T-\zeta)\eqno (7)
$$
where $\zeta ^2=r^2+z^2\tan^2 \! \psi$ ($z$ denotes height above the Galactic 
plane). Here the flattening parameter $\psi$ is introduced: $\cos \psi=q=c/a$, 
i.e. its cosine determines the shape of the 
halo $En$. The $En$ notation is related to $q$ as $q=1-n/10$. Following Sackett \& Gould 
(1993) we write the following expression for the estimate of the optical depth 
as a function of Galactic coordinates $l$ (longitude) and $b$ (latitude): 
$$
\tau(l,b)={\tan\psi\over \psi} {v_\infty^2 \over c^2}{1\over D}
\times
$$
$$
 \times \int\limits_0 
^{D} \! \! {(D-L)L \,dL\over (r_c^2+R_0^2)-(2R_0\cos l \cos b)L 
+(1+\sin^2b\tan^2\psi)L^2}
$$
$$\; \eqno(8)
$$ 
where we use $R_0=8.5$ kpc and $r_c=5$ kpc
(e.g.~Alcock et al. 1996). Now we integrate this equation and take $D=50$ kpc (for 
LMC), $D=63$ kpc (for SMC) and $D=770$ kpc for M31 (Binney \& Tremeine 1987). While
 Sackett \& Gould  
(1993) use values for $q$ starting with $q=0.4$ (shape E6), we will start with 
admittedly extreme value $q=0.2$ (shape E8) suggested by some theories, like the 
halo molecular clouds or the decaying dark matter (see the
discussion below).

There are several other lines of reasoning suggesting a high degree of halo 
flattening in spiral galaxies. One is for long time suspected (e.g.~Ninkovi\'c 
1985, 1991; Bahcall 1986; Binney et al. 1987) flattening of the Population II subsystem, 
which may be a consequence of 
the residual rotation, or more probably, global flattening of the 
gravitational potential created by dark matter. A detailed discussion of these questions was 
given by Binney et al. (1987), who attributed the flattening to a highly
anisotropic velocity dispersion tensor, under the assumption that the velocity
ellipsoid of halo objects near the Sun has the same shape as that of the extreme Population
II subsystem. Their calculation show that the parameter $q$ for the halo isodensity
contours has to be in the range $q=0.3 - 0.6$. Wyse \& Gilmore (1989) review
different arguments based on analyses of halo star counts for flattened halo and
suggest values $q=0.6$ as the optimal one. The same is, with additional data,
repeated in the extensive review of Gilmore, Wyse \& Kuijken (1989) where the
authors concluded that star count data definitely favor $0.6 < q < 0.8$ models.
One should note that widely used Bahcall-Soneira model of the Galaxy indicates
$q=0.8$, although it is realized that it may be too conservative, even for
stellar subsystem (Bahcall 1986). These results supersede older arguments
for spherical stellar component (e.g. Oort \& Plaut 1975).
As far as other galaxies are concerned, observational data are still extremely 
scarce, but it is indicative that very recently, the observations of the 
gravitational lens system B1600$+$434, 
consisting of two spiral galaxies (G1 and G2), where G2 is a barred one, 
suggest that it has $q^>_\sim 0.4$ (Koopmans, de Bruyn \& Jackson 1998). 
The flattening of the M31 halo was discussed and appropriate corrections
to the mass estimates were considered by Ninkovi\'c \& Petrovskaya (1992). 

As  pointed out by Sciama (1990), a further theoretical virtue of the 
halo flattening idea is connected with the Oort limit. Since the amount of the dark
matter per unit surface area of the disc is determined by the rotational velocity of 
the disc, any flattening of the vertical dark matter distribution must be compensated for by an increase in the density of dark matter in and near the galactic plane (i.e. in the disc).
Therefore, the amount of the dark matter near the Solar system, traditionally
associated with the Oort limit is reduced, and could be even brought down to zero (see also
Binney et al. 1987). This is appealing, since many recent results show incompatibility with the large quantities of local unseen matter (e.g.
Binney \& Merrifield 1998). Two different approaches yielded different values
for the  local density, $\rho_0$ near the Sun (the Oort limit): Kuijken \& Gilmore
(1989) found that $\rho_0=0.10\; {\rm M_\odot}\; {\rm pc^{-3}}$, while Bahcall (1984)
found $\rho_0=0.21\; {\rm M_\odot}\; {\rm pc^{-3}}$. However, it seems 
that recent {\it HIPPARCOS} measurements  favour lower value 
($\rho_0=0.11\pm 0.01 \; {\rm M_\odot}\; {\rm pc^{-3}}$) (e.g.~Kovalevsky 1998). This
value is not in disagreement with the assumption of maximal disc (Sackett 1997).

It is, also, necessary to consider the   behaviour of 
gas distributed in the halo. If the seminal idea of Bahcall \& Spitzer 
(1969) of extended gaseous haloes of normal galaxies producing narrow 
absorption features in the spectra of background objects is correct, as 
indicated by recent low-redshift measurements (Bergeron \& Boiss\'e 1991; 
Spinrad et al.~1993; Lanzetta et al.~1995, Chen et al.~1998), then the distribution 
of gas could tell us 
something about the shape of the gravitational potential. It is not a simple 
problem at all (e.g.~Barcons \& Fabian 1987; Pitts \& Tayler 1997), but 
some results are quite suggestive. In an important recent paper, Rauch \& 
Haehnelt (1995, hereafter RH95) have shown 
that for the most plausible values of Ly$\alpha$ cloud parameters, the 
conclusion that their axial ratio (thickness/transverse length) is less than 
0.25 is inescapable. This conclusion does not depend on the exact choice of 
model for Ly$\alpha$ clouds, and, if the observations quoted above are 
correctly interpreted, would mean that the gaseous haloes are also flattened by 
the same amount. One should keep in mind that such absorption studies 
probe only "a tip of an iceberg", since these objects are ionized to 
extremely high degree, and may as well contain dominant part of the baryonic 
density in the Eq.~(1). We shall return to discussion of these questions in the
Section 6.

\section{Optical depths for flattened haloes}

Bearing this in mind,  we solve the integral in the Eq.~(8)  and give estimate 
for $\tau$ in several cases of particular interest: 

\item {$\bullet$} Optical depth $\tau (l,b)$ in the parametric space, with the 
parameter $q$ fixed in steps of 0.2, i.e. $q=0.2$, $q=0.4$, $q=0.6$, $q=0.8$ 
and $q\approx 1$. 

\item {$\bullet$} Optical depth $\tau$ for different targets: LMC, SMC, M31 
and Galactic bulge (bar) in order to see what value of $q$ determines the 
optical depth that is closest to observed value in the corresponding survey. 

\noindent We hereby present several such three-dimensional 
plots. In  Figs.~1, 2 and 3, values of the optical depth $\tau$  
are plotted against 
galactic coordinates $l$ and $b$ for three chosen halo models, with
$q=0.2$, 0.6 and 1 (spherical halo), respectively. These axial ratios
were chosen as representatives of extremely flattened (like Pfenniger et al.~[1994]
or the DDM models discussed below), moderately flattened (like those favored 
by Bahcall-Soneira models or other dynamic discussions) and unflattened
models (usual approximation). One can use such plots (on the same  or  
smaller angular scales) in order to choose the observing direction where the 
optical depth reaches the maximal value. Such an example is shown in  Figs.~4
through 6, where we have plotted optical depth as function of coordinates 
$l=280.^\circ 5$ and $b=-32.^\circ 9$  of the Large Magellanic Cloud. 

\beginfigure{1}
\psfig{file=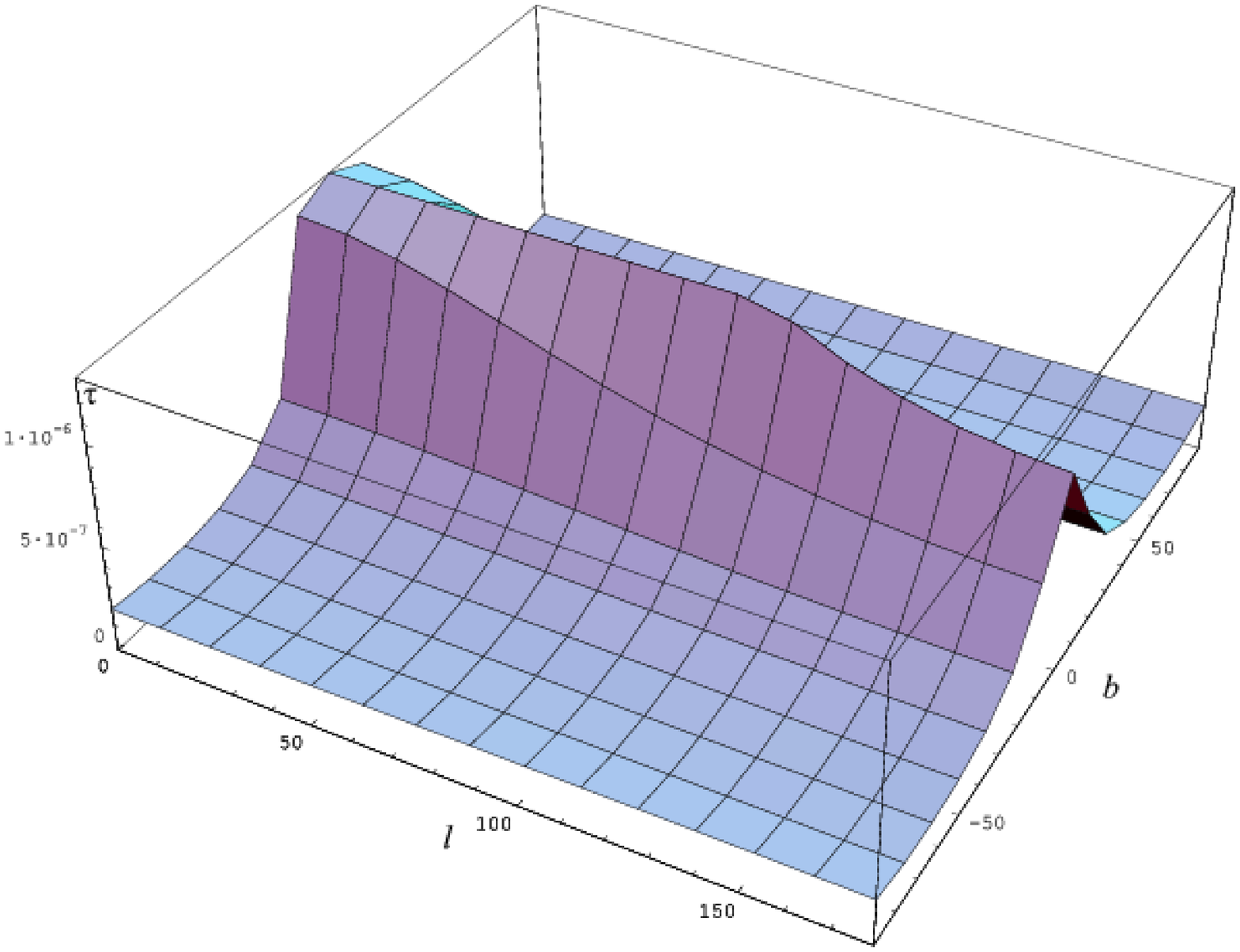,width=8cm}
\caption{{\bf Figure 1.} Optical depth for sources located at $D= 50$ kpc in a flattened model 
with $q=0.2$.}
\endfigure		

\beginfigure{2}
\psfig{file=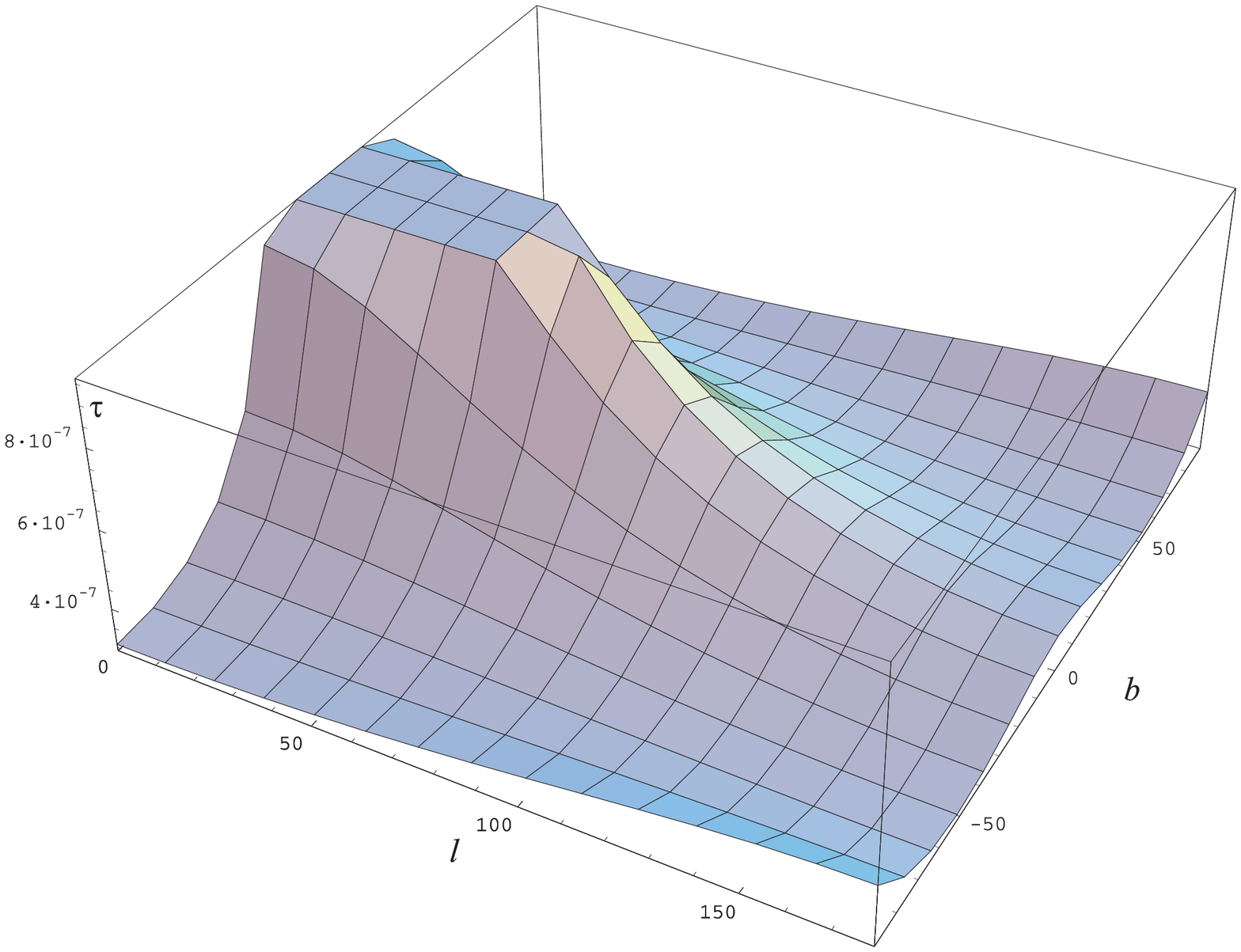,width=8cm}
\caption{{\bf Figure 2.} The same as in Fig.~1, except that model with intermediate flattening, $q=0.6$
is chosen.}
\endfigure	       %

\beginfigure{3}
\psfig{file=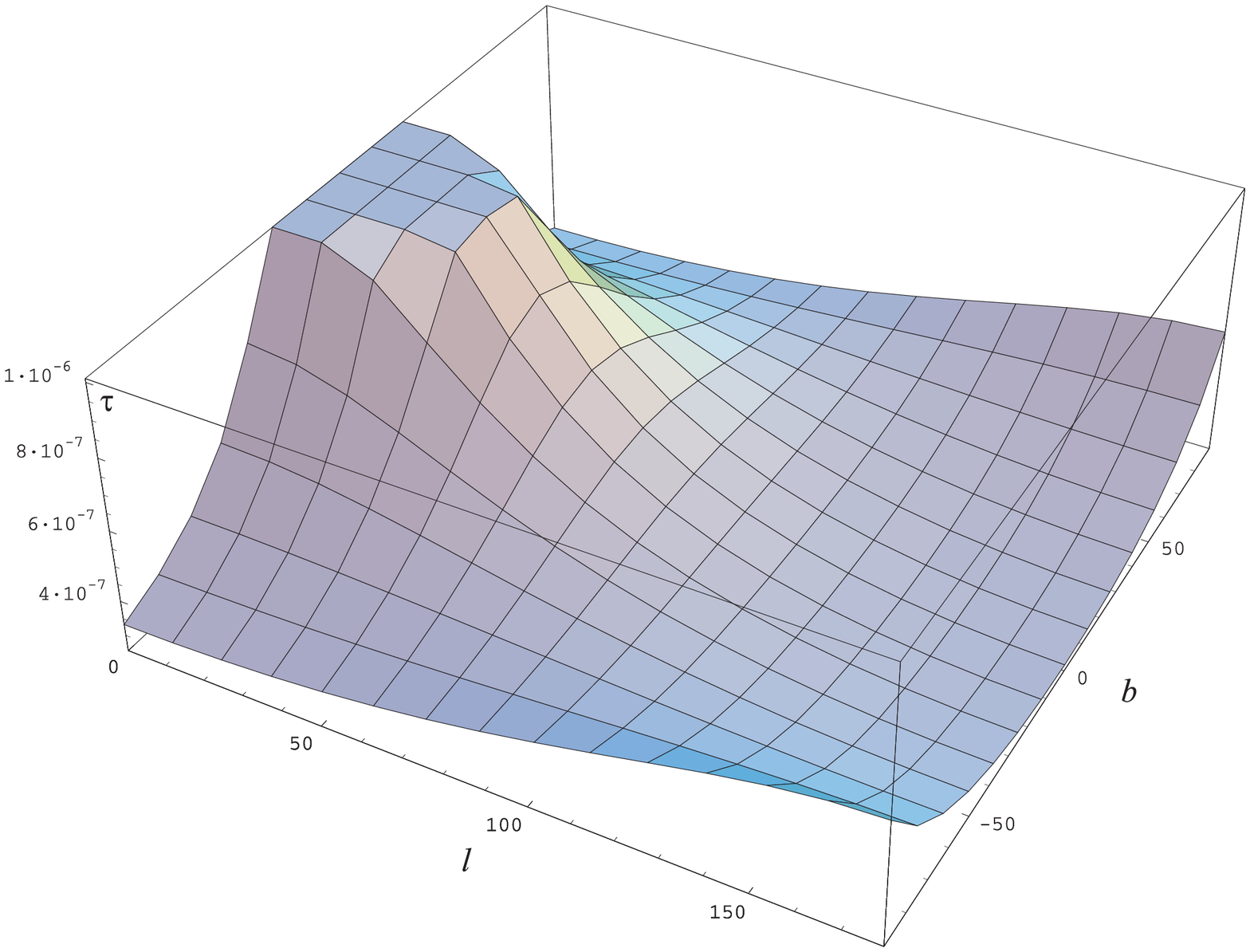,width=8cm}
\caption{{\bf Figure 3.} The same as in  Figs.~1 and 2, but for a spherical ($q=1$) case.}
\endfigure	    

After solving the integral in the Eq.~(8) for given values of the parameter  
$q$ we looked for the values that match the optical depth obtained in various 
surveys. We note that, in general, the best agreement can be attained with 
$0.2^<_\sim$ $q$ ${}^<_\sim 0.6$. Specifically:

\item {1.} For the case of the  LMC, that has been studied rather well, the 
measured value of the optical depth based upon the sample of 8 events is 
$\tau=2.9^{+1.4}_{-0.9}\times 10^{-7}$ (Alcock et al. 1997b) while we find that for 
$q=0.5$ we have $\tau\approx 3\times 10^{-7}$  (see Fig. 5). 

\item {2.} For the case of the SMC, that is studied less thoroughly, the 
optical depth is estimated as $\tau=1.5-3\times 10^{-7}$ (Alcock et al. 1997c).   Our 
results show that the model in the Eq.~(8) gives the value $\tau ^>_\sim 
4\times 10^{-7}$ for $q\approx 0.5$ and above. 

\item {3.} For the case of the galaxy M31 we found $\tau\approx 5\times 10^{-
6}$ which is an accordance with the estimates $5-10\times 10^{-6}$ (Crotts 
\& Tomaney 1996), under the assumption that $q^>_\sim 0.2$. 

\item{4.} Determining $\tau$ towards the Galactic centre is more 
complicated and we will not discuss it here. We only state that using the model in
the Eq.~(8) we can estimate the halo contribution to the ML rate towards
Galactic centre which is between $\tau\approx 5\times 10^{-7}$ ($q=0.2$) and 
 $\tau\approx 1.6\times 10^{-7}$ ($q\approx 1$); the estimated
range for the total optical depth towards Galactic centre is 
($\tau=3.9^{+1.8}_{-1.2}\times 10^{-6}$) (Alcock et al. 1997a). 
This discrepancy might be the consequence of the fact that the Galactic
disc is maximal: if this is so, then one can expect a higher optical depth 
towards the bulge due to disc lenses and 
lower values of optical depth towards targets in the halo, that is
measured, cf.~Table 1 (Sackett 1997). This nice illustration of the 
possible discrimination between various halo shapes, is presented in
 Figs.~7, 8 and 9 (the same cases as above).

\beginfigure{4}
\psfig{file=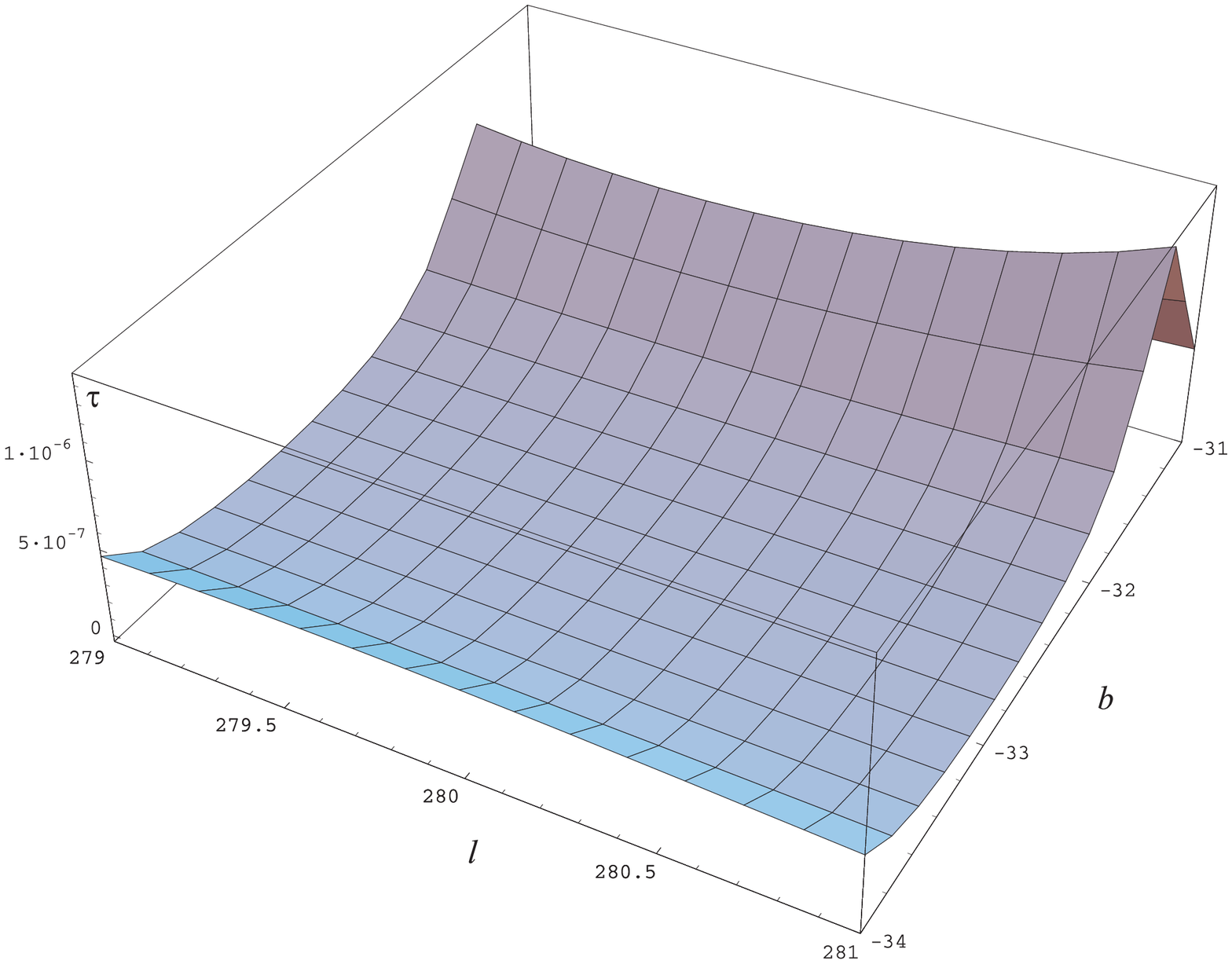,width=8cm}
\caption{{\bf Figure 4.} Optical depth toward LMC for very flattened ($q=0.2$) case.}
\endfigure	    

\beginfigure{5}
\psfig{file=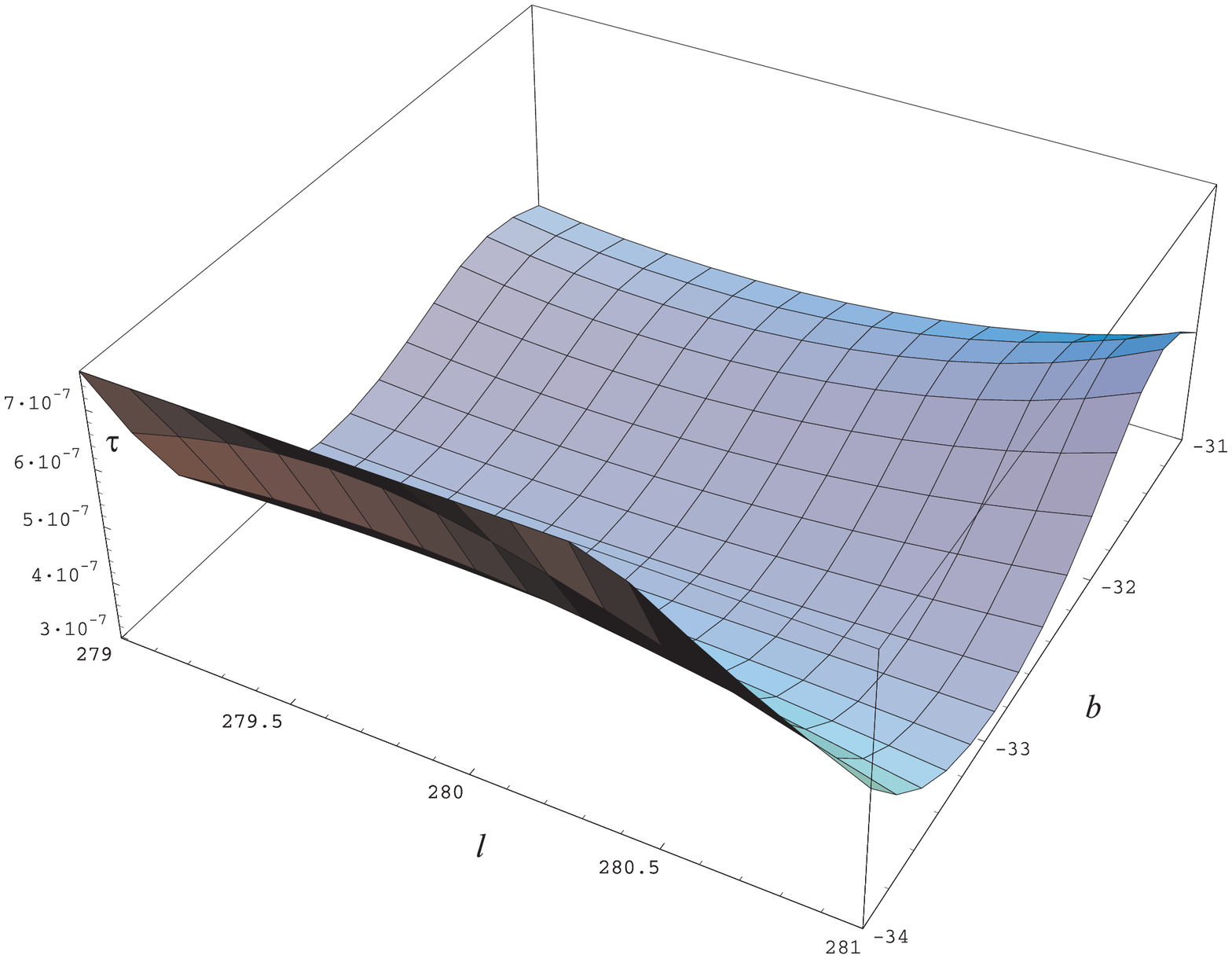,width=8cm}
\caption{{\bf Figure 5.} The same as in  Fig.~4, except that $q=0.6$ case is shown.}
\endfigure	    

\beginfigure{6}
\psfig{file=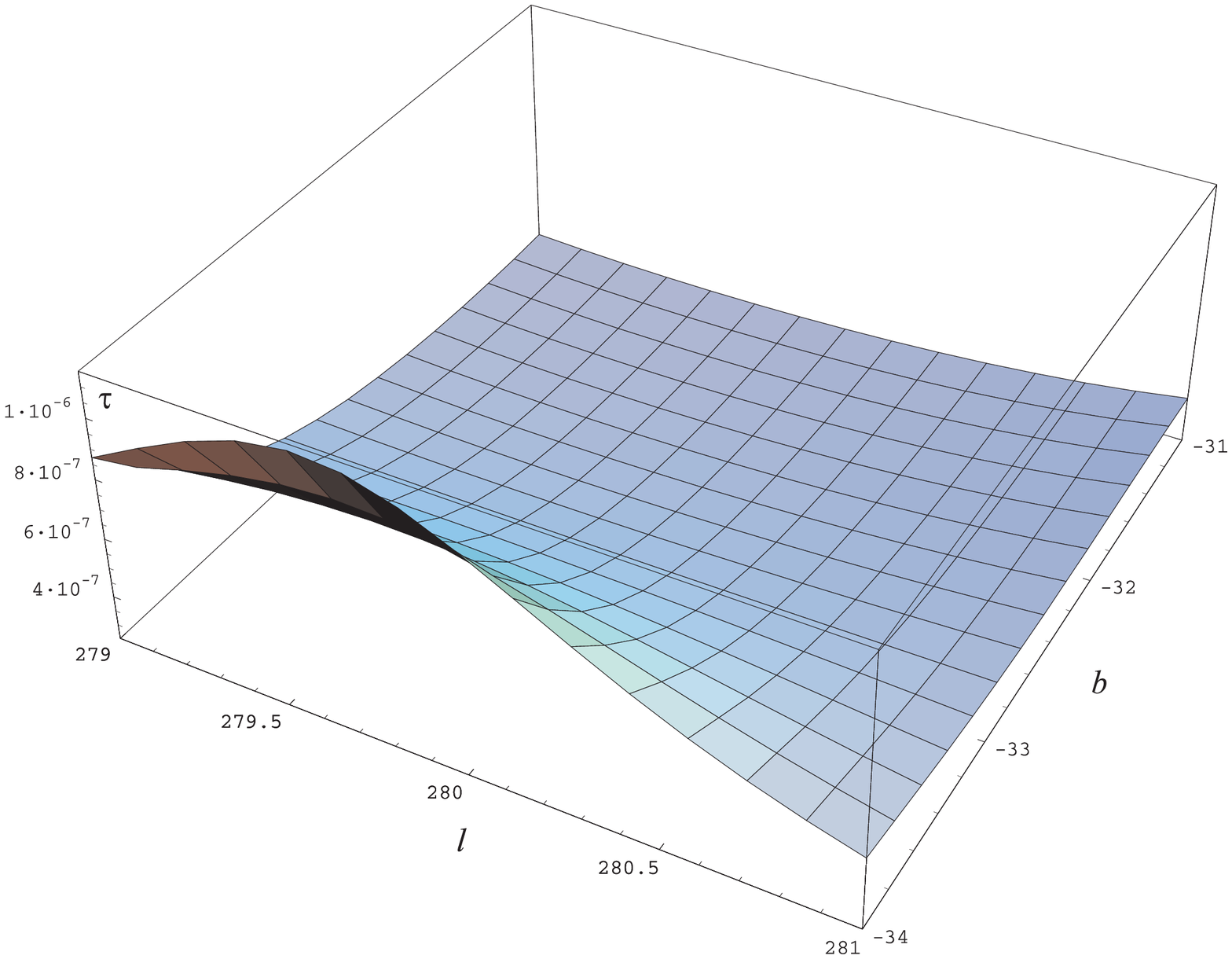,width=8cm}
\caption{{\bf Figure 6.} The same as in previous two Figures, but for spherical 
($q=1$) model.}
\endfigure	    

\noindent From these estimates and studies summarized in Table 1, 
it seems that both the spherical ($q\sim 1$) and extremely flattened ($q \simeq 0.2$) 
dark halo can be ruled out with high significance. 
The optimal value for $q$ is the intermediate one
$q \simeq 0.6$. Recent research shows that it is not uncommon 
case with spiral galaxies, not only polar-ring ones 
(Sackett \& Sparke 1990; Sackett et al.~1994; Olling 1995). 
This has several far-reaching consequences, like implausibility
of molecular halo or DDM models, which, in the simplest forms,
predict strong flattening of dynamically important halo component. 

We also present ratios of optical depth toward LMC and SMC as functions
of the flattening parameter $q$ in Table 2. This is significant, since variation in 
$\tau_{\rm LMC} /\tau_{\rm SMC}$ is claimed to sufficiently clearly 
discriminate between various models of flattening (Sackett \& Gould 1993). Our
results are in agreement with the more detailed discussion of this
problem by Frieman \& Scoccimaro (1994).

The necessity of having  many more lines of sight for microlensing survey besides LMC, SMC,
 Galactic bulge and Andromeda
galaxy, led several research groups to consider globular clusters as ML targets
 (Rhoads \& Malhotra 1998, Jezter et al.~1998,
Gyuk \& Holder 1998). The original idea came from Paczynski (1994)
  who
proposed 47 Tuc (monitored by OGLE collaboration) and M22.  Gyuk \&
 Holder (1998) and Jezter et al.~(1998) 
composed  lists of appropriate clusters for microlensing survey, with corresponding optical depths,
stating that by using globular clusters it would be possible to distinguish the
flattened models of the galactic halo much easier, and it would allow better determination of halo
structure parameters, such as the power-low index, or the core radius. Special advantage of that
method is avoiding self-lensing events present in a SMC survey, for example, and fluctuations in halo
density due to clumps of stars in tidal tails and intervening dwarf galaxies  like recently discovered
Sagittarius galaxy.
 
We briefly note that very recently  the EROS collaboration (Derue et al. 1999)
found that in the four directions towards the Galactic spiral arms
the average value for the optical depth is equal 
$\bar \tau=0.38^{+0.53}_{-0.15}\times 10^{-6}$. If one compares values obtained 
for their four targets
and the predicted values of $\tau$ from Fig. 2, one could  again conclude that
the halo is moderately flattened, i.e., $q\sim 0.5-0.6$.

With globular clusters distributed throughout the Mil\-ky Way's halo, this method will further 
improve on the number of ML events
and make it possible to constrain Milky Way's dark halo shape and content more precisely.   

It is of foremost importance to continue theoretical studies of these
optical depths, even if we are not in position to find other possible sources
of background light for microlensing experiments. Further improvement in 
number of events will enable precise measuring of abovementioned optical depths, 
and using their ratios to constrain possible halo shapes. 

\beginfigure{7}
\psfig{file=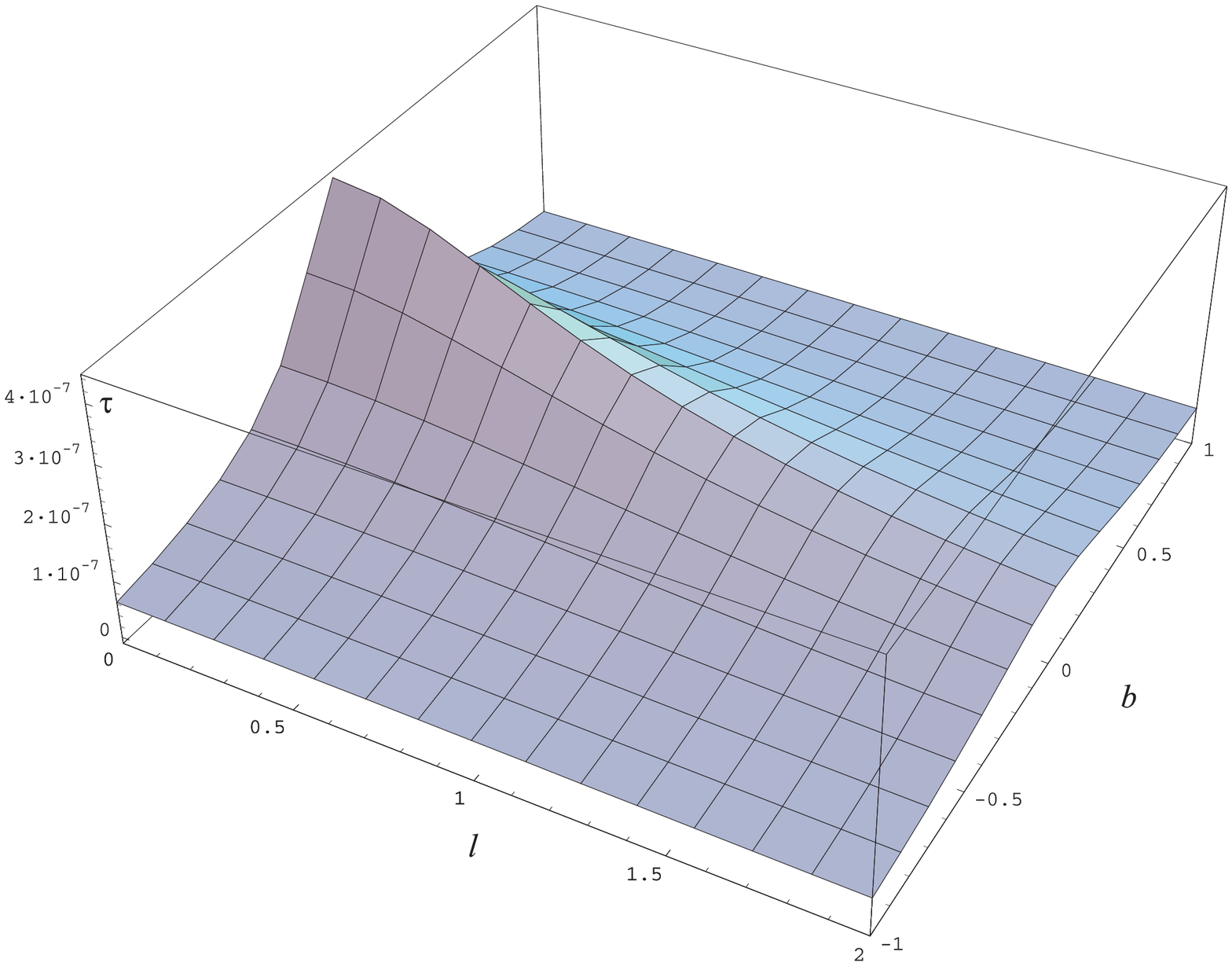,width=8cm}
\caption{{\bf Figure 7.} The halo contribution to the microlensing optical depth toward the 
Galactic Centre for $q=0.2$.}
\endfigure		

\beginfigure{8}
\psfig{file=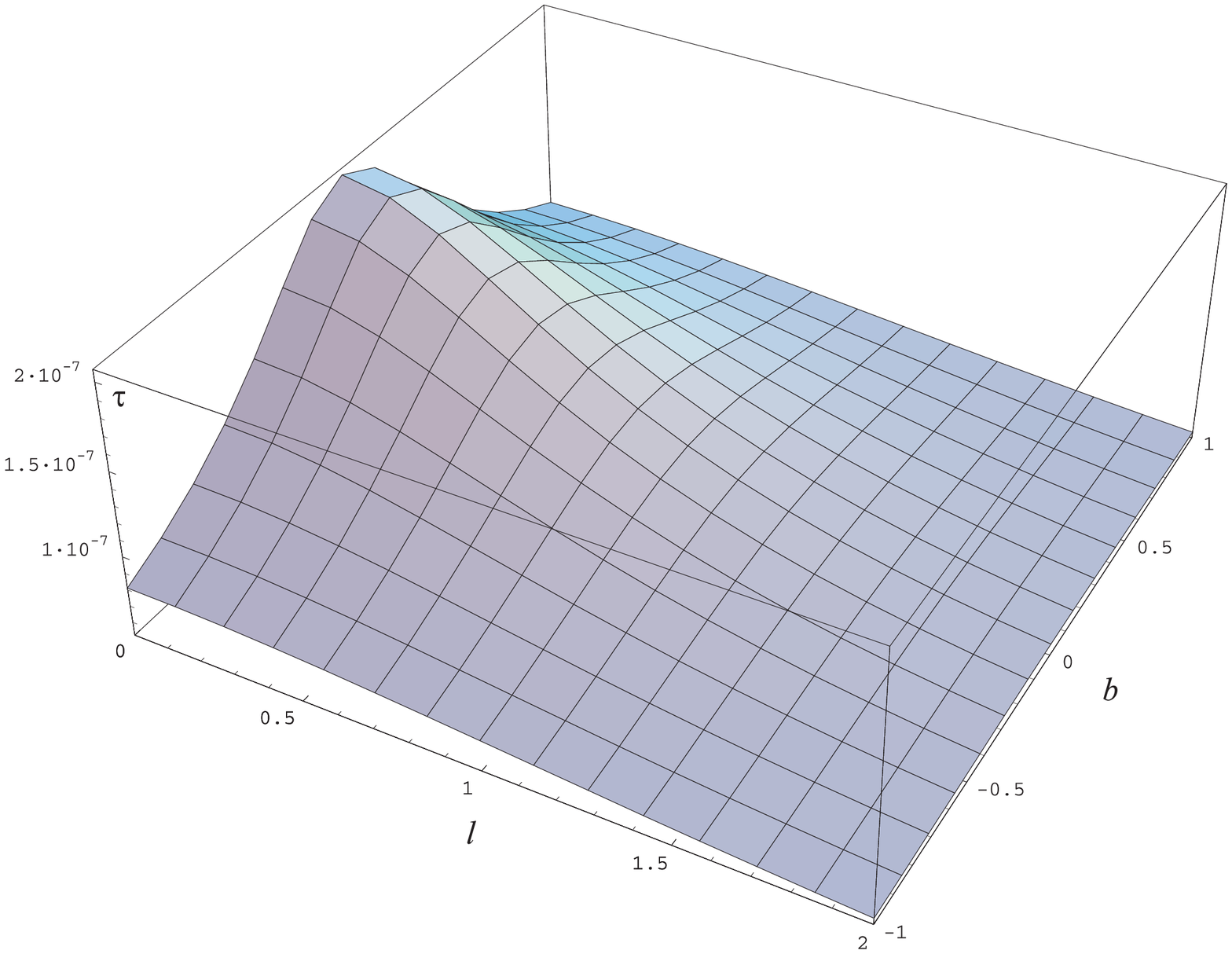,width=8cm}
\caption{{\bf Figure 8.} The same as in Fig.~7, for moderately flattened halo ($q=0.6$).}
\endfigure	       

\beginfigure{9}
\psfig{file=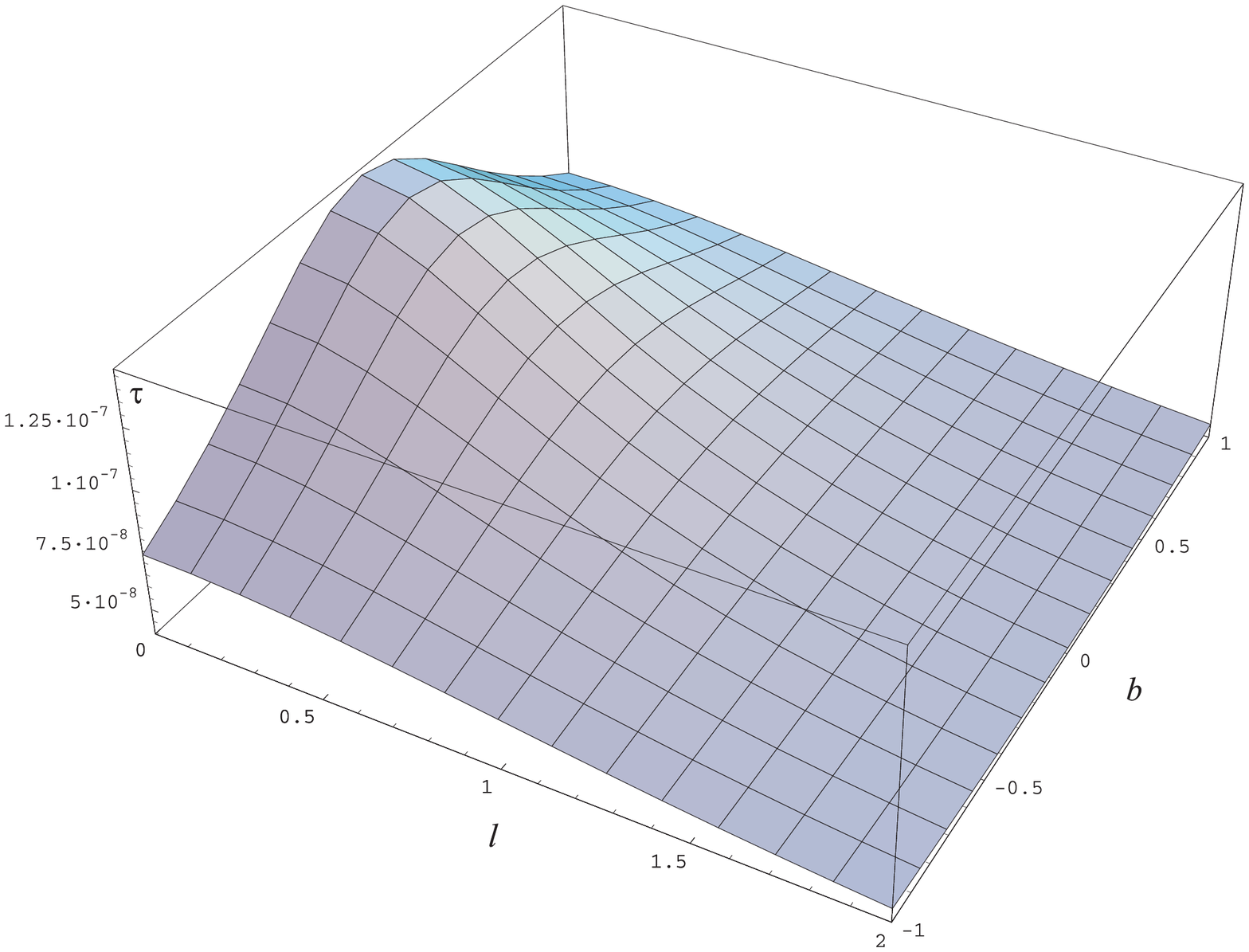,width=8cm}
\caption{{\bf Figure 9.} The same as in previous two Figures, for the spherical case.}
\endfigure	       

\begintable*{2}
\caption{{\bf Table 2.}  $\tau_{\rm SMC}/\tau_{\rm LMC}$ ratio for different values of the parameter 
$q$. Optical depths, $\tau$, are  expressed in units of $10^{-7}$. }
\halign{%
\rm#\hfil&\qquad\rm#\hfil&\qquad\rm\hfil#&\qquad\rm\hfil
#&\qquad\rm\hfil#&\qquad\rm\hfil#&\qquad\rm#\hfil
&\qquad\rm\hfil#&\qquad\rm#\hfil&\qquad\hfil\rm#\cr
$q$ & $\langle \tau_{\rm LMC}\rangle$ & $\langle \tau_{\rm SMC}\rangle$ &

${\langle \tau_{\rm
	SMC}\rangle\over \langle \tau_{\rm LMC}\rangle}$\cr
	&&&\cr
 0.2 & 3.8554 & 5.2082 & 1.3509\cr
 0.6 & 4.0733 & 5.6973 & 1.3987\cr
 1.0 & 4.2184 & 5.6670 & 1.3434\cr
}
\endtable
\section{Total mass of MACHOs and cosmological density parameter}

The total mass of a MACHO halo of an $L_\ast$ galaxy (as typified by the Milky Way)
in a model characterized by the Eq. (7) is
$$
\eqalign{M(q,R_T) & = 3.648 \times 10^{-12} \frac{\sqrt{1-q^2}}{q \, \arccos q} \times \cr
 & \times \int\limits_0^{R_T} \! \int\limits_0^{R_T q} \! \frac{r \, dz dr}{r_c^2 + r^2 + z^2 
\frac{\sqrt{1-q^2}}{q}} \, {\rm M_\odot},} \eqno(9)
$$
where all lengths are in cm, and $r_c$ has a fixed value. We shall briefly discuss
on the variation of core radius below.

\beginfigure{10}
\psfig{file=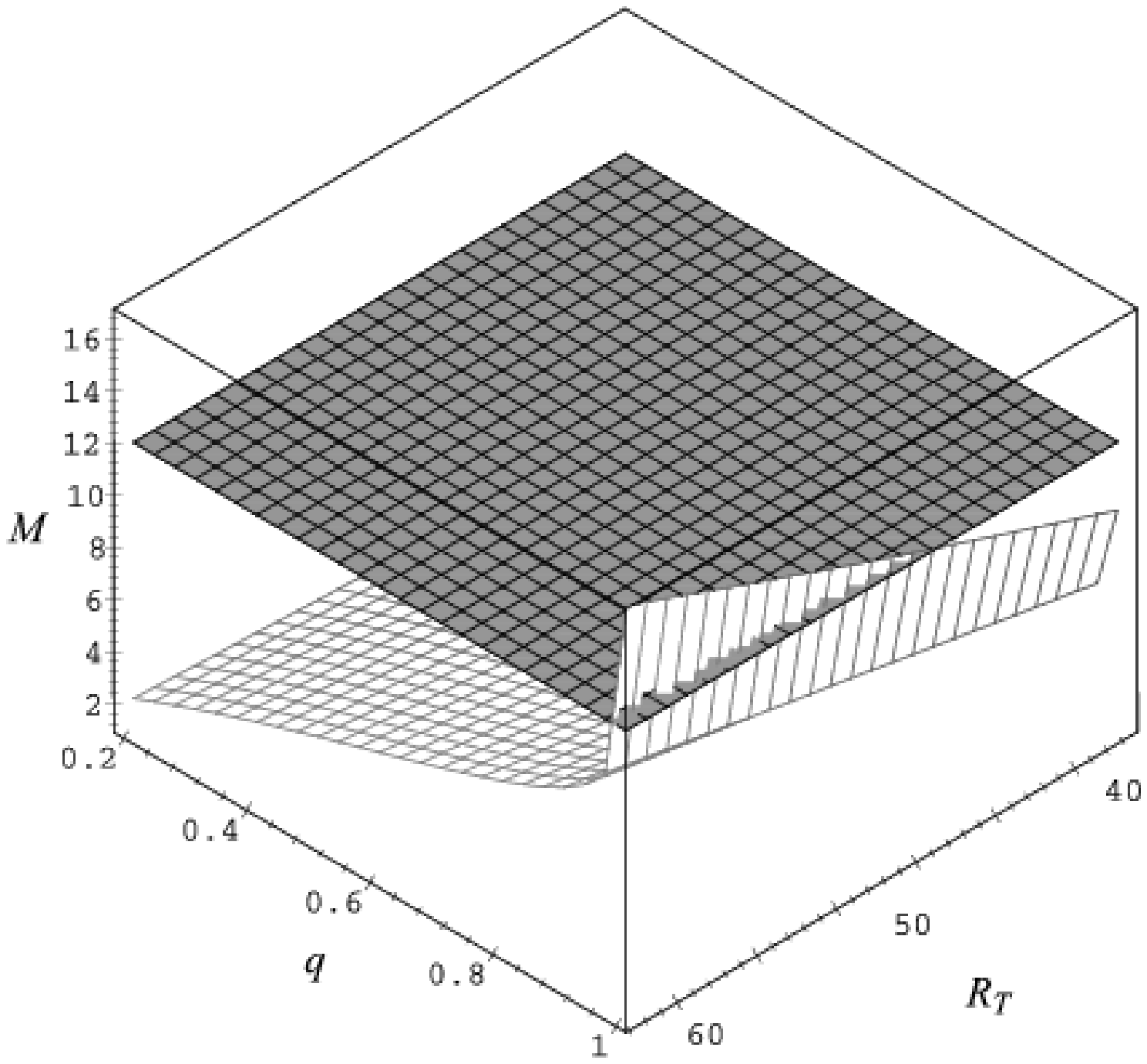,width=8cm}
\caption{{\bf Figure 10.} The mass of the Galaxy in 
units of $10^{11}{\rm M_\odot}$ as a function of the truncation radius,
 $R_T$ and the flattening parameter, $q$. Canonical value for the core radius, $r_c =5$ kpc is used in this plot.}
\endfigure

The total mass of an $L_\ast$ galaxy is shown in  Fig. 10 for $q$ varying between 
0.2 and 1, and a truncation radius $R_T$ between 40 and 60 kpc. Horizontal plane represents the dynamical value of the total mass inferred from satellite studies within much larger radius of $\sim 230$ kpc by Kulessa \& Lynden-Bell (1992). The choice of
interval for the truncation radius is relevant not only because it incorporates the "canonical"
value of 50 kpc for the size of MACHO haloes (FFG; Alcock et al. 1996, 1997a), 
which is reasonable from the
point of view of empirical detection of microlensing events toward Magellanic Clouds;
another, entirely theoretical, argument is that the cooling times for protogalactic halo gas in this region ($R \leq 50$ kpc) are an order 
of magnitude shorter than the dynamical time (Rees \& Ostriker 1977; White \& Rees 1978),
thus making a collapse of baryonic structures a likely outcome. We shall extend somewhat
the discussion of the relevant physics in the Sec.~7. As we shall see from the plots in 
this and subsequent figures, further increase in $R_T$ leads to huge masses of MACHO
haloes, which are unacceptable from the point of view of BBNS, unless the fraction of 
galaxies containing MACHO haloes similar to the one of the Milky Way is, for some
quite mysterious reason, very small. 

We also note that MACHOs are incapable of explaining galactic dynamics on large
scale, and in order to explain dynamical estimates of the Milky Way mass
based on satellite systems (Kulessa \& Lynden-Bell 1992; Zaritsky et al. 1997)
dark matter in form of either invisible gas or non-baryons has to be invoked. As
we shall see below, if one chooses to accept the BBNS constraints for the
baryonic cosmological density $\Omega_B$, non-baryonic dark matter in the
Milky Way halo (and haloes of all normal galaxies) has to be invoked. On the 
other hand, the presence of non-baryonic dark matter within MACHO halo itself
(i.e.~at galactocentric distances up to $\sim 50$ kpc) has the effect of relaxing
bounds on the MACHO halo, and thus is, in principle, subject to empirical
verification through detailed comparison of observed optical depths with those
predicted by the density profile in the Eq.~(7).

\beginfigure{11}
\psfig{file=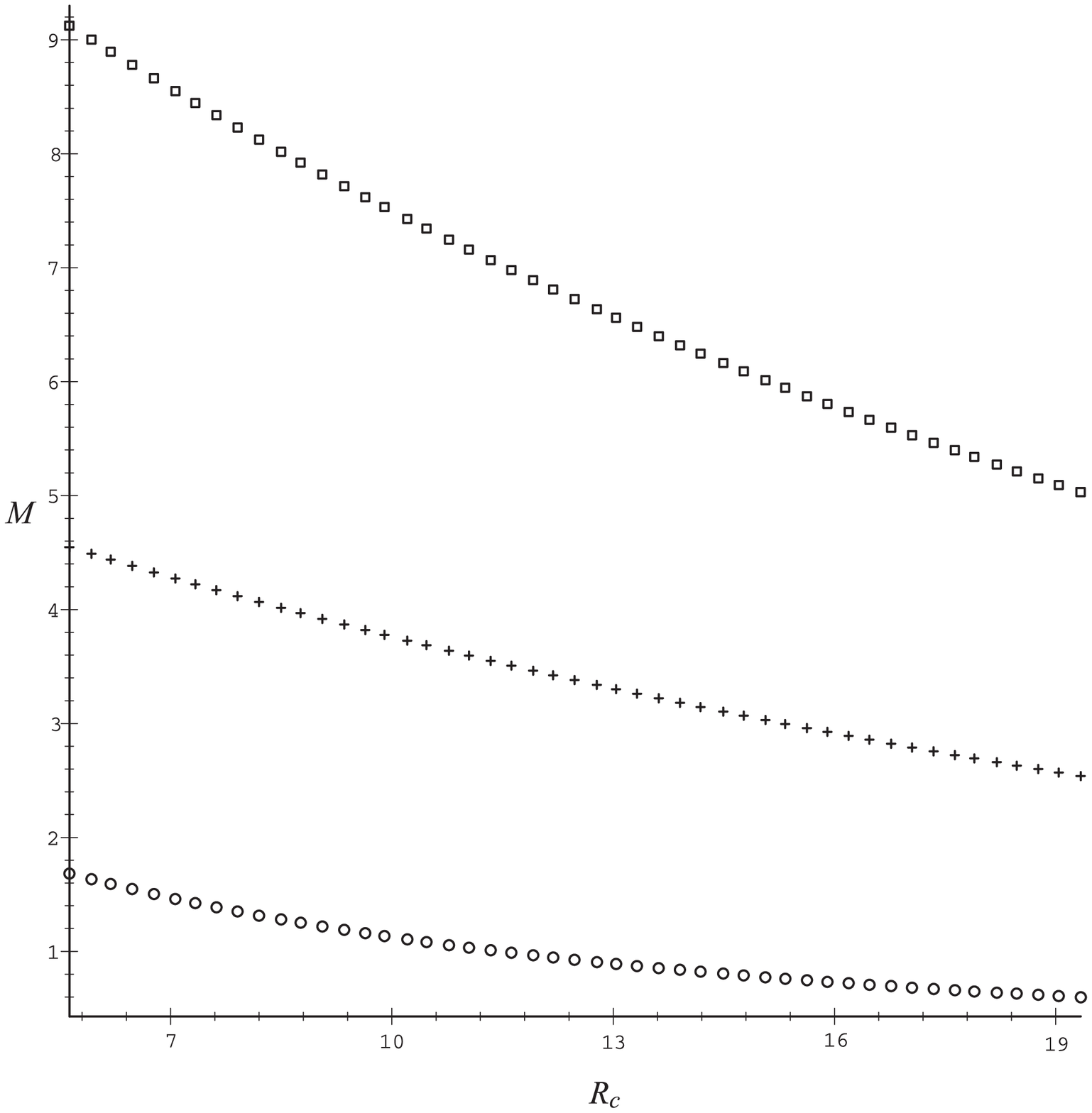,width=8cm}
\caption{{\bf Figure 11.} Masses of MACHO haloes (in units of $10^{11}\; {\rm M_\odot}$)
for three different choices of the flattening parameter $q$, as a function of the core radius $r_c$ for
a fixed truncation radius of $R_T =50$ kpc. Square points represent $q=1$ case, circles $q=0.2$ extremely
flattened case, and crosses the intermediate one ($q=0.6$).}
\endfigure	   

As far as other components of the total cosmological density in baryons $\Omega_B$
are concerned, we adopt the Persic \& Salucci (1992) estimate for the mass of visible baryons
(i.e.~stars, ISM and gas in rich clusters) given by the Eq.~(4). 
Further contribution is expected to come from the intergalactic and/or "invisible"
galactic gas. At later epochs, this is what FHP call "warm gas around galaxies and
small groups". Since these are presumably the same objects as those recently detected as
the dominant fraction of the low-redshift Ly$\alpha$ forest, we shall denote this
contribution as $\Omega_{\rm Ly\alpha}$ in further discussion. 

There are basically two ways in which one can discuss relationship between Ly$\alpha$
clouds and baryonic dark matter: (i) direct comparison of their cosmological
density in various epochs (as in FFG), 
and (ii) we can consider transformation of 
high-$z$ Ly$\alpha$ clouds into present day MACHOs. We proceed with (i), and shall
return to the topic (ii) in the next section.

In order to translate individual galactic masses, as in the Eq.~(9) into
global cosmological density parameter $\Omega_{\rm MACHO}$, it is
necessary to perform integration over the luminosity function (LF), with some 
assumptions. Beside universality of the luminosity function, we have to
assume that there is no diffuse, intergalactic population of MACHO-like objects.
This is not just a formal statement -- it puts obvious constraints on epoch of 
formation of such objects and their degree of clustering. It is natural to
speculate that, due to dynamical effects, some MACHOs will be ejected from the
halo during galactic history, thus creating such an intergalactic population,
with its own particular contribution to the value of $\Omega_B$. In this sense, our
present picture is not completely self-consistent, since it neglects this 
intergalactic population of collapsed objects (expression "MACHO" is, obviously,
inadequate here). In the course of future work, we hope to quantify this assumption in detail
and, especially, demonstrate implications for high-density regions (e.g.~rich 
clusters), where "sharing" of the BDM among galaxies may have crucial
influence upon its evolutionary history, and result in observable 
peculiarities (e.g.~White \& Fabian 1995). 

Cosmological density parameter in such MACHOs residing in haloes
of typical luminous galaxies can be written as 
$$
\Omega_{\rm MACHO} = \frac{1}{\rho_{\rm crit}}
 \int\limits_{L_{\rm min}}^{L_{\rm max}} \! \! M(L) 
\varphi(L)\, dL, \eqno(10)
$$
where $\rho_{\rm crit}$
 is the critical density of
Friedmann-Robertson-Walker universes,
$L_{\rm min}$ and $L_{\rm max}$ are the minimal and maximal luminosity
of galaxies possessing such MACHO haloes, respectively, 
and $\varphi(L)$ is the universal 
galaxy LF (Schechter 1976; Binggeli, Sandage \& Tamman 1988; Willmer 1997). 
We use the LF in Schecther's form
$$
\varphi (L) ={\varphi_\ast}   \left  ({L\over L_\ast} \right    )^{-\gamma}
\!\! \exp  \left(  -{L\over L_\ast} \right), \eqno(11)
$$
and alternative mass form can be found in Nulsen \& Fabian (1997). LF parameters
are chosen to be (Willmer 1997)
$$
\varphi_\ast = 2.5\times 10^{-2}\, h^3\; {\rm Mpc}^{-3}, \eqno(12)
$$
and 
$$
\gamma = 1.27. \eqno(13)
$$
The influence of their variation on our results is discussed in detail below. 
The standard Schechter luminosity is chosen to be $L_\ast = 1.0 \times 10^{10}\, e^{\pm 0.23}\, h^{-2} \, L_\odot$, i.e.
corresponding to the absolute B-band magnitude of $M_\ast = - 19.2$
(Willmer 1997). The Hubble parameter $h$ is chosen to be $h=0.5$ in all calculations, except where otherwise mentioned.

\beginfigure{12}
\psfig{file=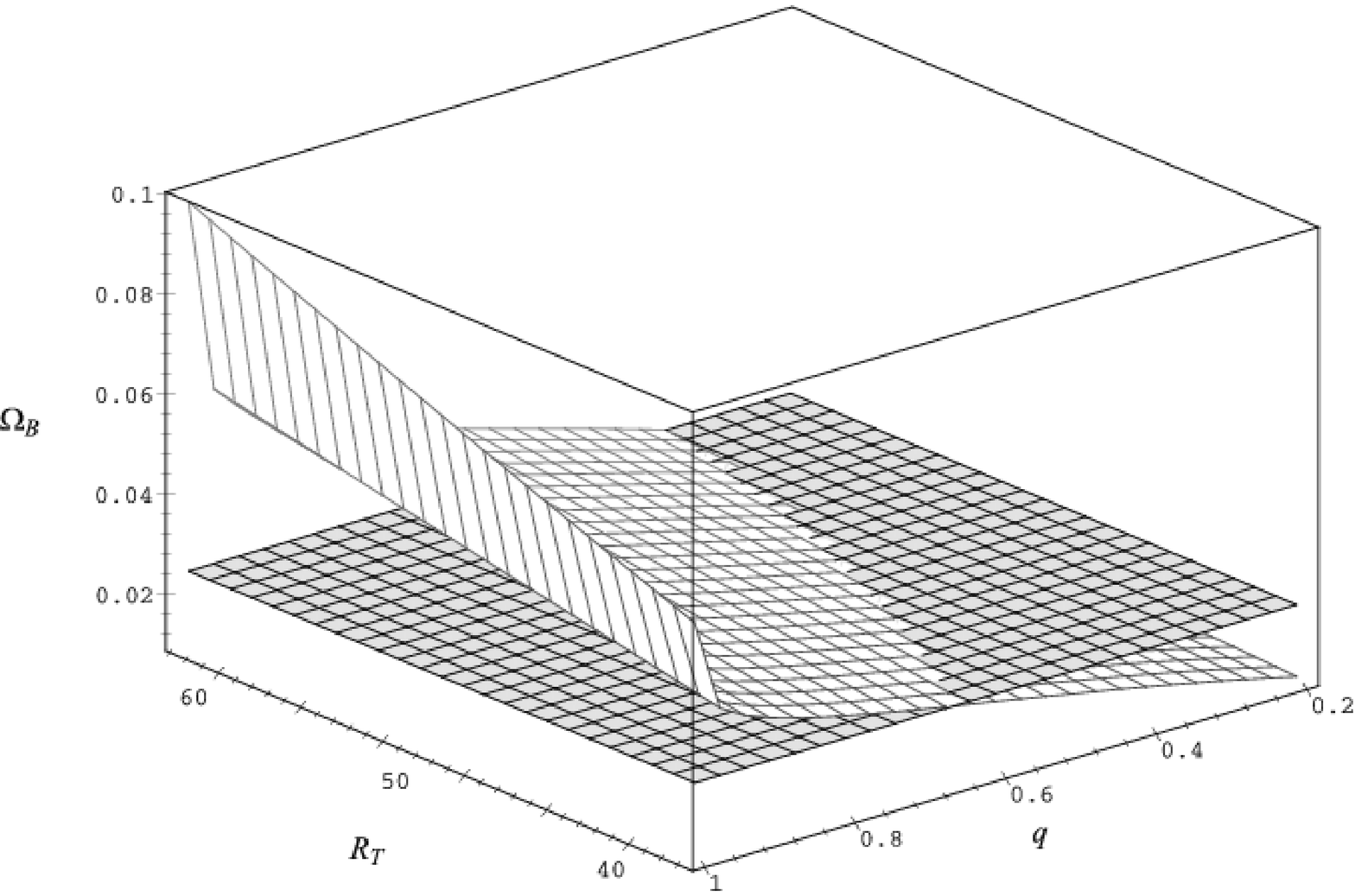,width=8cm}
\caption{{\bf Figure 12.} The cosmological density parameter $\Omega$ in MACHOs plus
visible baryons as a function of the truncation radius,
 $R_T$ and the flattening parameter, $q$. Canonical value for the core radius,
$r_c =5$ kpc is used in this plot, and $h=0.5$. Lower bound from 
the primordial nucleosynthesis is also shown, for comparison.}
\endfigure
 
We note from the Eqs. (9) and (11) that a unique value of $\Omega_{\rm MACHO}$
corresponds to each pair of values $(q, R_T)$. The distribution of possible values of
this cosmological density parameter is shown by the 3-D plot in  Fig. 12 for $r_c =5$
kpc and $h=0.5$.

We have used fiducial value for the core radius $r_c = 5$ kpc. Variation of this 
quantity in the usual range 5$-$8 kpc (Binney \& Tremaine 1987) 
causes changes in our results of 
$\delta M / M = \delta \Omega / \Omega \leq 12$ per cent (see Fig. 11). In addition, we have also investigated somewhat
unorthodox value $r_c = 20$ kpc (but, see Gerhard 1999). This is motivated by some recent indications that 
the Milky Way rotation curve may be satisfactorily explained by nearly homogeneous
dark matter distribution within a few solar circles (Ninkovi\'c, private 
communication); see also Frieman \& Scoccimarro (1994). 
Also, such a large core-radius would be completely in accord 
with suggestion, originating with N-body simulations (Cole \& Lacey 1996), that 
the density profile becomes significantly flatter then the isothermal one in
inner halo regions (becoming simultaneously steeper than $r^{-2}$ in outermost
regions). The influence of varying core radius on the mass of a fiducial halo
with our model profile is shown in  Fig.~11. 

\beginfigure{13}
\psfig{file=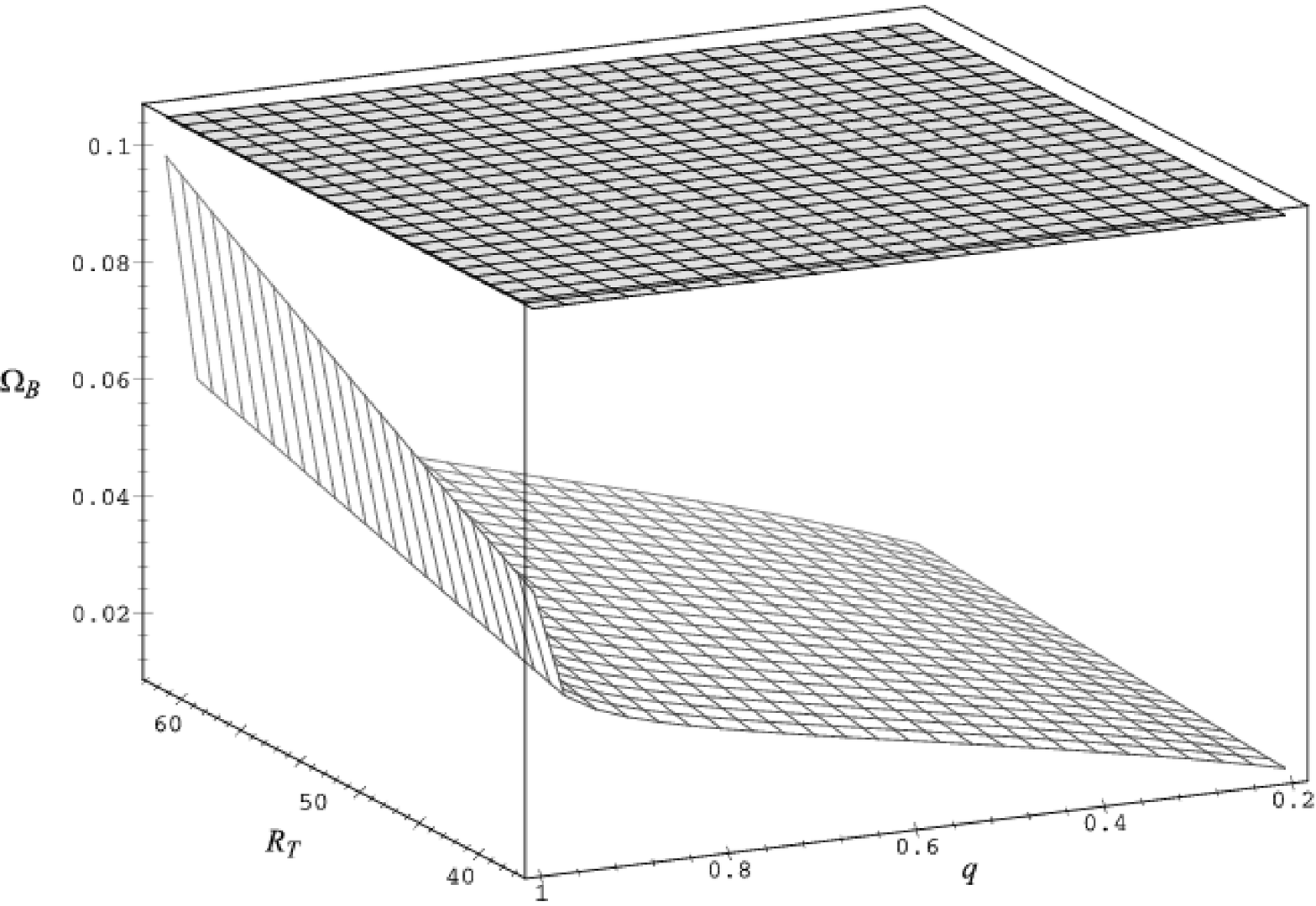,width=8cm}
\caption{{\bf Figure 13.} The same as in  Fig.~12, except that the upper 
nucleosynthetic bound is shown (for the same choice of parameters).. It seems 
clear that we need a significant gaseous baryonic component in the present day universe, 
quite in accord with the low-$z$ Ly$\alpha$ forest observations. }
\endfigure

Variation of other parameters also does not remedy the high value of the total
mass in MACHOs. For example, $v_\infty$ is only bound from below, by the IAU
value of Galactic rotation at the solar circle of 220 km s$^{-1}$ which was
used in these calculations. If, as indicated, Milky Way rotation curve rises all the way 
to $\sim 3 R_0$ (Ninkovi\'c, private communication), $v_\infty$ can be as high as 280 km s$^{-1}$ 
(Frieman \& Scoccimarro 1994), and the mass $M(q,R_T)$ would be increased for
a factor $\approx 1.62$ with corresponding increase in $\Omega_{\rm MACHO}$, 
which, taking into account the bounds in the Eq.~(1), is not insignificant.

In  Figures 12 and 13, we have shown the total MACHO + visible cosmological density 
vs. the constraints emerging from the BBNS for $h=0.5$. 
We notice that MACHOs within 50 kpc are certainly capable of solving the problem of missing baryons
resulting from comparison of the Eqs.~(2) and (4). Again, exceptions are very
flattened haloes with $q \simeq 0.2$, which make them still less appealing possibilities.
On the other hand, if higher nucleosynthesis estimates of $\Omega_B$, for example 
$\Omega_B \approx 0.077$ (Burles \& Tytler 1998), are reconfirmed by 
impending observations, there seems to be no alternatives to discarding little flattened
values $q \geq 0.8$ either.

One should always keep in mind that there is no physical reason for assumption that 
$R_T$ is close to the canonical value of 50 kpc; rather, it is just an empirical 
convenience, at least for the time being. Caution suggests to take these values 
(i.e. $R_T$ and the corresponding masses) as lower limits only, with consequences
that flattening looks even more appealing as a way to reduce $\Omega_{\rm MACHO}$.
If MACHO haloes extend to anything similar to the extent of dynamical haloes
inferred, for example, by Zaritsky et al.~(1997), then rejecting of anything with
$q > 0.6$ is unavoidable. 
On the other hand, it is possible that our reliance on the Occam's razor is misleading,
and only some fraction $f$ of the mass distribution creating potential
responsible for the rotational curve is in form of MACHOs. The rest  $1-f$ must
then be in the form of non-baryonic dark matter, if we wish to remedy the high
$\Omega_B$ problem, which leads to a degeneracy, where flattened full-MACHO halo may
contain the same amount of mass as non-flattened realistic halo with $f<1$. On
the other hand, optical depth estimates and ratios discussed in the Sec.~4 
would still be dependent only on the MACHO fraction, and therefore optical depths
are expected to be reduced by the factor $f$ and their ratios to be unaffected.
This offers a further opportunity for improved microlensing statistics, which should
be able to easily discriminate between values of $f$ close to unity and any other 
significantly smaller value (at present, as visible from the Table 1 and comparison with
Figs.~4 through 6, we can only claim $f > 0.1$ with reasonable certainty).

\section{Gaseous content of galaxies and flattening}

The discovery that a large fraction of low-redshift Ly$\alpha$ forest is associated
with normal luminous galaxies (Spinrad et al.\ 1993; 
Lanzetta et al.~1995; Chen et al.~1998) presents a 
further difficulty for the total baryonic census, as  recognized by FHP.
This means that at least some debris from the galaxy formation epoch remained in
the gaseous state till relatively late epochs; the question whether this gas
(discovered up to huge galactocentric distances, with maximal absorption radius
for $L_\ast$ galaxies being $\sim 300$ kpc) was partially recycled through some
galactic stellar population is unimportant in this respect. Possible contribution
to the baryonic budget is enormous; it is the largest (albeit the most uncertain)
entry in the list of FHP. In fact, its magnitude is such that the BBNS constraints
are seriously jeopardized by a direct extension of column-density statistics
to the total mass contained along all lines of sight, an insight which prompted
RH95 to suggest a significant flattening of these gaseous
structures. 

In that work, it is shown that constraints following from the general formula for the cosmological density of Ly$\alpha$ systems
$$
\Omega_{{\rm Ly}\alpha} = \frac{\mu m_{\rm H} H_0}{c\rho_{\rm crit}} \!\!
\int\limits_{N_{\rm min}}^{N_{\rm max}} \! x^{-1} N f(N)\, dN, \eqno(15)
$$
(where $N$ is the {\it neutral} hydrogen column density spanning the
interval between $N_{\rm min}$ and $N_{\rm max}$, $x$ neutral gas fraction, and 
$f(N)$ the neutral hydrogen column density distribution) coupled with the BBNS 
bounds leads to inevitable conclusion of global flattening, if sizes (or coherence)
lengths obtained from double line-of-sight analyses are taken seriously. Typical 
values of $\Omega_{{\rm Ly}\alpha} \simeq 0.04$ are obtained for typical 
sizes of 100 kpc, in spherical case, from several of their simple models, which is, 
obviously, quite high. 
If large coherence sizes inferred from double lines of sight (e.g.~Dinshaw et
al.~1995) are characteristic (and they are in general accord with the huge sizes of
{\it galactic} gaseous haloes obtained by Chen et al. [1998]) the cosmological density
is even higher. The way out is to assume that axial ratio of these structures
(without entering the question of their physical origin and location) is small,
and for the most conservative of their models, RH95 obtain $q_{{\rm Ly}\alpha} \leq
0.1$. It is interesting that they suggest clumping of the neutral content
as an alternative way of decreasing the total mass, a frequent 
suggestion which has not 
been fully investigated to date (e.g. Mo \& Miralda-Escud\'e 1996). 

Conclusions of RH95 are, it should be reemphasized, essentially 
independent of the true nature and location of 
the Ly$\alpha$ forest clouds. They are valid for both inter- and intragalactic
types of absorbers. But the discovery of large population of halo absorbers
at $z \leq 1$ prompts us to ask whether aspect ratio of absorbers in RH95 can,
in fact, be interpreted as the flattening parameter of gaseous haloes. In addition, 
not only the fact that low- and intermediate-$z$ absorbing clouds preferentially
lie in galactic haloes, without noticeable morphological segregation (Yahata et al.
1998), but also the fact that the covering factor of such haloes was found to be close to
unity everywhere within the absorbing radius (Chen et al. 1998), suggests that
gaseous haloes should be flattened at late epochs. It should be noted that 
flattening of Ly$\alpha$ absorption systems was much earlier proposed, 
for different reasons, by Barcons \& Fabian (1987) and Milgrom (1988).

Unfortunately, exact knowledge of the baryonic content of the Ly$\alpha$ 
absorbing clouds requires certain knowledge on their ionization structure,
which is still very elusive. The value of metagalactic ionizing background, 
which is the only always operating ionizing source, is still painfully uncertain
 even in the local universe
and at low redshift, and the more so at high-$z$ 
(Bajtlik, Duncan \& Ostriker 1988; Kulkarni \& Fall 1993;
Vogel et al. 1995; Donahue, Aldering \& Stocke 1995). The presence of internal
ionizing sources, inferred in some local intergalactic clouds (Donahue et
al. 1995; Bland-Hawthorn et al. 1995) is also quite uncertain. Finally, geometric
properties of ensembles of clouds, i.e. their clumpiness and global flattening,
are still only speculative.

The comparison of FFG with the mass estimate of the Ly$\alpha$ forest of Weinberg 
et al. (1997) is interesting, especially in the view of possible absence of mixing
between the two types of unseen baryonic matter. Weinberg et al. (1997) value,
quoted by FFG,
$$
\Omega_{{\rm Ly}\alpha} = 0.02 \, h^{-\frac{3}{2}},  \eqno(16)
$$
is not at all especially high when considered within a "family" of closure
fractions obtained for $\Omega_{{\rm Ly}\alpha}$  

Higher values are required for high-$z$ intergalactic Ly$\alpha$ forest by many
models, e.g. Bi \& Davidsen (1997) suggest $\Omega_{\rm Ly \alpha} = 0.025 \, h^{-2}$,
adopting the values of metagalactic ionizing flux from Haardt \& Madau (1996).
Alternatively, one may wish to reduce the value of $\Omega_{\rm Ly \alpha}$, but
at a price of having significantly different ionizing background (since models
are able only of constraining choices for $\Omega^2 /J_{\rm UV}$). Furthermore,
any increase in the cosmological bias would tip the scales toward larger
contribution of gas in comparison to the visible matter assembled in stars
and "normal" luminous galaxies.

In our opinion, the association of significant fraction (if not all) low-$z$
Ly$\alpha$ forest with galactic haloes does give additional credence to FFG
conjecture about MACHOs being distinct baryon reservoir from Ly$\alpha$ forest.
At least this is so after some particular epoch, which we shall denote
by $z_{\rm bd}$, when bulk of today's MACHOs was formed out of gas-rich
protogalactic fragments, and which may be called the epoch of baryonic decoupling. 

Gaseous density profile and global shape should, in principle, follow
profile and shape of the underlying dark matter distribution, which is
supposed to be dynamically dominant (Nulsen 1986). This is the necessary link
between arguments concerning flattening of gaseous structures around galaxies
and flattening of MACHO halo. Of course, exact amount of flattening (as well 
as the exact density profile) would be different because dark matter is,
in contradistinction to gaseous matter, assumed to be dissipationless (no matter
whether it is mainly MACHO or non-baryonic elementary particle), but
the difference is not expected to be very large on the scales of $\sim 50$ kpc.
If there is a significant infall from the halo to the disc, the constraints
could be much tighter. In this respect, it is important to mention that an
argument of this type was advanced by Sancisi \& van Albada (1987), namely,
that recent accretion of gas found in the optical plane 
at outer fringes of spiral galaxies already implies flattening of the dynamical
halo in these regions. 

Conclusions of FFG actually receive multifold support from considerations of 
low-redshift Ly$\alpha$ forest:

\item {1.} Probable overestimate of the $\Omega_{\rm Ly \alpha}$ in N-body
simulations and other models of unclustered population of Ly$\alpha$ clouds
makes it easier to accommodate large $\Omega_{\rm MACHO}$.

\item {2.} The very fact that we perceived much stronger Ly$\alpha$ absorption at recent epochs than expected on the basis of naive extrapolation supports the conclusion that Ly$\alpha$ 
clouds are rather decoupled from star-formation histories (at least after some fiducial epoch $z_d$), so the discussion in the Sec.~3 of FFG is justified. 

In addition, recent observational indications that Milky Way still possesses
extended gaseous halo with densities $\sim 10^{-4}$ cm$^{-3}$ at galactocentric
distances $\sim 50$ kpc (i.e.~similar to those discussed with respect to the
MACHO halo) underlie the necessity of having large fraction of dark baryons in
gaseous form at present day (Weiner \& Williams 1996). This further narrows available
range for $\Omega_{\rm MACHO}$, and suggests that all possibilities to reduce it
should be explored, flattening being the simplest one of them and most in line with
the principle of economy of hypotheses. 

\beginfigure{14}
\psfig{file=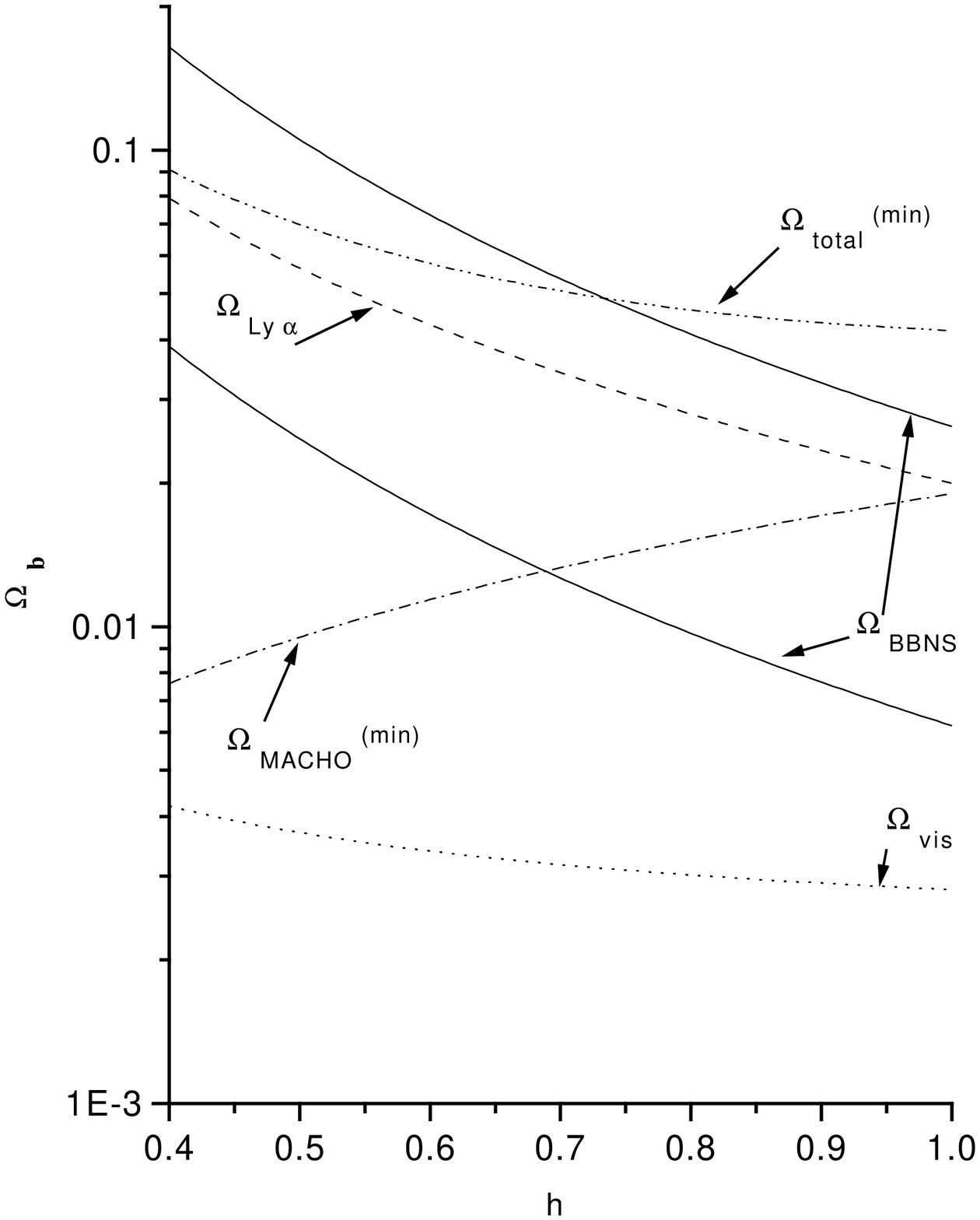,width=8cm}
\caption{{\bf Figure 14.} Various components of the baryonic cosmological density (visible, 
Ly$\alpha$ absorbing gas and MACHOs) compared with the BBNS bounds for varying Hubble
parameter $h$. In this Figure, the minimal mass of the MACHO halo within 50 kpc in our
discussion, corresponding to $q=0.2$, is taken. The sum of all three main baryonic 
components is shown as the dash-double-dotted line labelled with $\Omega_{\rm total}^{\rm (min)}$.}
\endfigure

\beginfigure{15}
\psfig{file=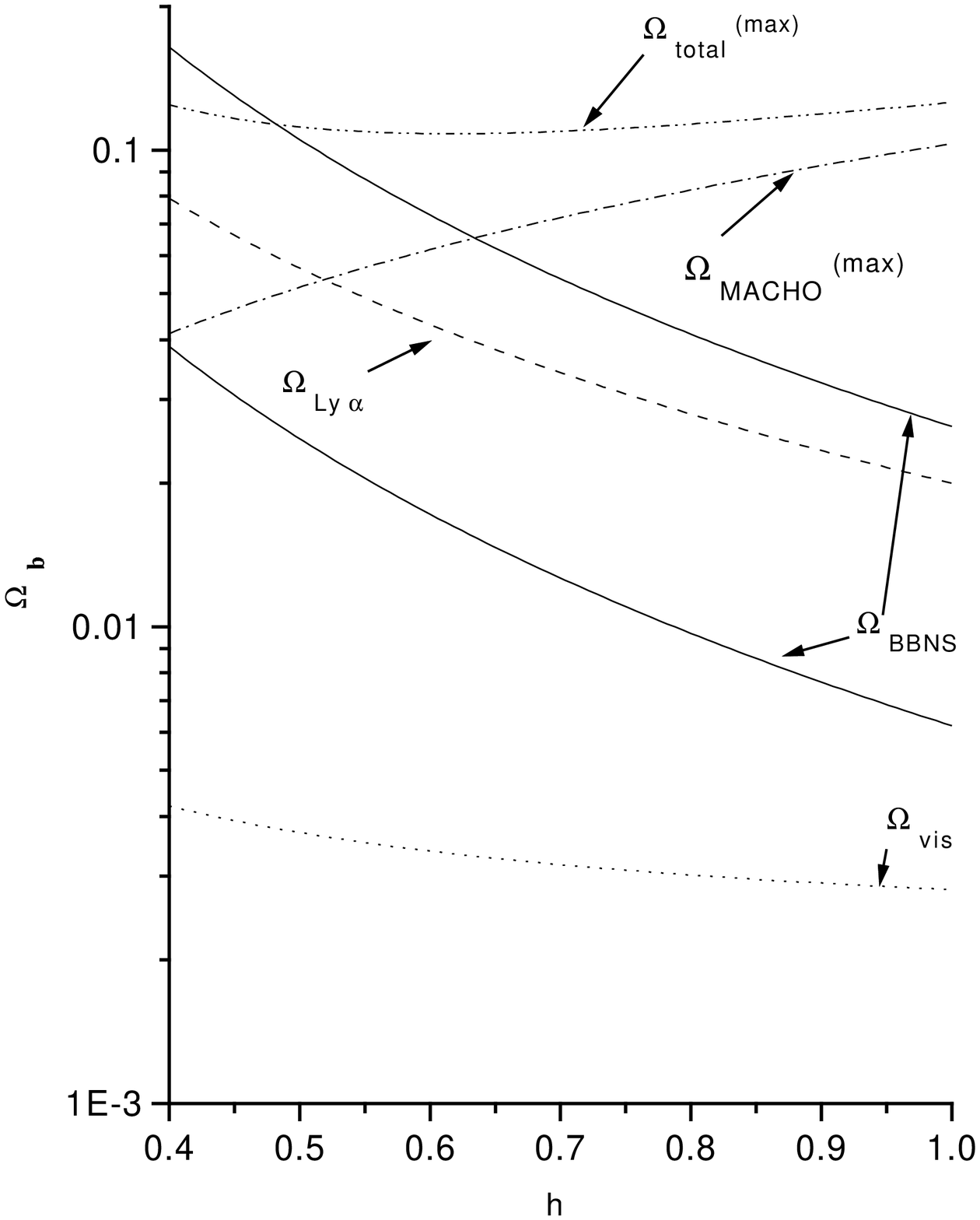,width=8cm}
\caption{{\bf Figure 15.} The same as in  Fig.~14, for the maximized MACHO cosmological
density (non-flattened $q=1$ case).}
\endfigure

Various baryonic components are represented in the $\Omega -h$ diagrams in Figs.~14 and
15. If the Ly$\alpha$ mass estimated by Weinberg et al.~(1997) is correct, and
the assumption of FFG about essential decoupling of the baryonic contents of
Ly$\alpha$ forest and MACHOs after some initial period, high values
of $h \geq 0.8$ seem to be highly implausible for both flattened and unflattened MACHO haloes. For spherical haloes, we ran into troubles for almost all values of $h$. It is marginally acceptable for $h=0.5$, but it is inconsistent with any higher values (again, we should keep
in mind that there is no physical reason for the truncation of MACHO halo at the LMC distance, and many arguments that dynamical haloes extend much further). Flattened haloes, on the other
hand, are quite securely within the BBNS margin for $h \simeq 0.5$.

\section{Discussion}

Mass considerations could, in principle, lead us to an important clue in
solving the puzzle of the fate of the halo gas. All scenarios of hierarchical
structure formation (e.g. White \& Rees 1978; Navarro \& White 1994) 
have massive dark haloes in place by $z \sim 2 - 3$,
which form gravitational potential wells accreting diffuse baryonic matter (Mo \&
Miralda-Escud\'e 1996). It is almost certain that at high and intermediate redshift
exactly this gas dominates cosmic baryonic budget (RH95; FHP). 
Its subsequent history is still shrouded in mystery, but the fact that 
estimates of the cosmological density fraction in MACHOs in both FFG and present
work point out that at some point in the galactic history the transition between
the halo gas (or its significant fraction) and collapsed objects, like the present
day MACHOs occurred. 

As the {\it phase transition of baryonic dark matter} we denote the process of 
transformation of pregalactic/proto- galactic gas into MACHOs. It seems obvious
that transition from the diffuse (gas) to the collapsed (MACHO) phase of the
BDM must have occurred at some point in the history of the universe. 
This process, whatever 
form and in whichever epoch it took place, is of the crucial importance for 
understanding the evolution of baryonic content of the universe. Parenthetically,
this is return to the authentic physical meaning of the concept of phase of
matter, since Ly$\alpha$ clouds and MACHOs do possess different symmetry properties.

This transition from gaseous to collapsed phase can proceed in several ways. 
While exact behavior is still too difficult to investigate in detail,
since it depends on the initial dynamical and chemical conditions, we 
can sketch several possibilities.  

An early transition probably occurs through Population III stars, and in
this scenario, MACHOs are stellar remnants (Carr 1994; FFG). 
Only detailed chemical modelling can show whether this is viable. 
Apart from possible overproduction of metals, a further problem with
this picture is the necessity for the Pop III initial mass function (IMF)
 to be significantly
different from that observed today in the ISM. This is not only to 
avoid conflicts with results of deep searches (Richstone et al. 1992;
Flynn, Gould \& Bahcall
1996), but also to avoid problems with massive stellar black holes (disruption of
stellar discs, excessive X-ray emission due to accretion, etc.). 

Otherwise, it can proceed via cold gas, passing through a phase similar
to that depicted in models of Pfenniger, Combes \& Martinet (1994), Pfenniger \& Combes (1994), Gerhard \& Silk
(1996) or Walker (1998) and Walker \& Wardle (1998). 
It is not clear how the IMF in such cases 
can be restricted to 
low-luminosity objects. On the other, it may be significant that Pfenninger et al.
 (1994) models do predict extreme flattening of baryonic haloes (or thick discs),
composed of fractal distribution of small, sub-Jeans mass cloudlets (Pfenniger
\& Combes 1994). 

Finally, gas may form MACHOs in an early {\it cooling flow}, similar to those
we perceive today in rich clusters and around isolated giant elliptical 
galaxies (Fabian, Nulsen \& Canizares 1982; Sarazin 1988; Fabian et al. 1986;
Nulsen, Johnstone \& Fabian 1987). 
This is scenario of Nulsen \& Fabian (1995; 1997). It has the 
advantage that it creates conditions favorable to low-luminosity objects
{\it in a natural way} (Sarazin \& O'Connell 1983; Sarazin 1986; 
Ferland, Fabian \& Johnstone 1994; 
Fabian \& Nulsen 1994). Further virtue of this picture is that it 
presents continuity with
cooling of hot, virial haloes recently invoked in order to explain metal and
Lyman-limit absorption systems (Mo 1994; Mo \& Miralda-Escud\'e 1996). 

As Nulsen \& Fabian (1997)  point out, cooling flow scenario
seems to be capable of solving the fundamental puzzle of origin of 
morphological difference between galaxies. This is even more important in
view of very recent QSO absorption results indicating that there is no
significant difference between absorption properties of 
low- and intermediate-redshift early- and late-type galaxies (Yahata et al.
1998). Since Ly$\alpha$ absorption naturally samples larger galactocentric
distances than optical surveys which produced all existing morphological  
classifications, it follows that outer regions, where the dynamical timescale 
is longer, are similar for all haloes of comparable total (i.e. baryonic + nonbaryonic)
mass. Since thermal instabilities, cooling and infall are necessary
consequences of an early virialization (e.g. White \& Rees 1978), 
one does find present-day cooling 
flows with their large mass deposition, a viable model for the evolution of
gaseous haloes. A parcel of infalling cooled gas, as pointed out 
by Binney (1995) and Mo \& Miralda-Escud\'e (1996), can have only two 
fates: it may fall into the disc and be supported by angular momentum,
effectively becoming part of the galactic ISM, or it can experience sufficient
cooling before reaching the disc, or the centre of the halo, and -- possibly
passing through transient molecular phase -- create stars and/or brown
dwarves. These two paths are likely to produce entirely different global BDM
structures; the former would create (and help to sustain) spiral galaxies,
and the latter would apply to triaxial systems. The detailed models along these
lines, extending the view of Nulsen \& Fabian (1997), may 
present a significant step forward in explaining the morphological
differences between galaxies.

In any case, it is our hope that future detailed modelling, coupled
with observational breakthroughs in both microlensing and halo absorption
systems will answer the question on the empirical value of the baryonic
fraction of gaseous haloes, which can be written as
$$
f_g = \frac{M_{\rm MACHO} + M_{\rm gas} + M_{\rm vis}}{M_{\rm tot}}. \eqno(17)
$$
This parameter is crucial for the theories describing post-virialization cooling and early 
infall of gas. Mo \& Miralda-Escud\'e (1996) choose $f_g =0.05$, value that is in
general agreement with orthodox assumptions of the BBNS, but otherwise remains a free 
parameter of the model. Large cosmological constant, for example, like
recently popular models with $\Omega_\Lambda \sim 0.6$ (e.g. Schmidt et al.
1998), although 
obviously not influencing the cooling of protogalactic gas, will manifest itself
indirectly through a significant increase in $f_g$.

Pietz et al. (1998) show that X-ray emission from the inner Milky Way
gaseous halo is best reproduced in a flattened model with $q \sim 0.3$,
although other flattened models give better agreement with the data than
the spherical models.

The flattened shape of the dark halo is not only important in connection 
with the BDM, but also with non-baryonic
dark matter such as massive neutrinos. Decaying dark matter theory 
(Sciama 1993),
which is based on the assumption that the major
constituent of  the dark mass in haloes of spiral galaxies is a massive decaying 
neutrino, with the mass of $\sim 30$ eV
and  lifetime $\sim 2 \times 10^{23}$ s, 
{\it requires} that the haloes are extremely flattened -- the halo of the Galaxy is flattened so 
the axial ratio is $q=0.2$ (Sciama 1990, 1997). 
This is a consequence of the effort to reconcile the observed 
electron densities obtained from the pulsar dispersion measure 
data (Nordgren, Cordes \& Terzian 1992) 
according to which $n_e\sim 0.03$ cm$^{-3}$. DDM theory would
 give the value  of $n_e\sim 0.017$ under the assumption of spherical halo. Agreement is thus obtained with significant flattening of the 
halo that reduces the scale height by the factor $\sim 4$ to $\sim 2$ kpc.
One can notice that this flattening could be achieved by extending the mass 
models of the Galaxy by Dehnen \& Binney (1998). However, recently, an attempt 
has been made to show that in the case of the Galaxy the axis 
ratio $q$ is  $q=0.75\pm 0.25$ thus ruling out cold molecular gas and decaying 
massive neutrino as viable
dark matter candidates (Olling \& Merrifield 1998).  It is somewhat beyond the scope
of the present paper, and we only notice that the aforementioned result with this, 
rather high, value of the parameter $q$ is attained if the
galactocentric distance is $R_0=7\pm 1$ kpc, which is a rather unorthodox value. On the
other hand, microlensing optical depths indicate that "standard" versions of
these theories, requiring $q \simeq 0.2$ are no longer viable. This conclusion can
be avoided if MACHOs are dynamically insignificant within 50 kpc.
However, it can be shown  that one can easily
accommodate a much larger value of $q$, i.e. $q\sim 0.6$, into the DDM theory without
significant changes of the theory's fundamental parameters such as mass and
lifetime of the decaying neutrino 
(Samurovi\'c \& \'Cirkovi\'c 1999).

\section{Conclusions}

On the basis of still scarce empirical data, and still undeveloped and
unsophisticated theoretical models, one can, however, draw some
important inferences regarding the global shape of haloes of spiral
galaxies, taking the Milky Way as a prototype $L \sim L_\ast$ galaxy
at zero-redshift. It seems that a whole array of different arguments
point to an oblate gravitating dark halo, which, in turn, causes flattening
of other types of haloes, like stellar and gaseous (Ly$\alpha$-absorbing) haloes.
Our conclusions are thus summarized as follows:

\item {1.} The set of measured optical depths for microlensing, 
although still statistically incomplete, strongly indicates moderately
flattened haloes.

\item {2.} We reconfirm the conclusion of FFG that $\Omega_{\rm MACHO}$ is very high
in all plausible cases, specifically, 
$$\Omega_{\rm MACHO} / \Omega_B= 0.1 - 1.$$
For spherical or little-flattened haloes, it is, in fact,
unpleasantly close to $\Omega_B$, as obtained from the primordial nucleosynthesis 
and flattening seems to be the simplest remedy for this  situation.

\item {3.} Review of relevant literature reveals a multitude of arguments for 
flattened haloes of the Milky Way and other spiral galaxies. We find the arguments
based on the total mass of Ly$\alpha$ forest clouds especially convincing, in conjunction 
with strong arguments for association of low and intermediate-redshift Ly$\alpha$ forest
with normal galaxies.

\item {4.} We find the FFG conclusion that baryons in MACHOs and Ly$\alpha$ forest are
essentially decoupled is strengthened on the basis of low-$z$ absorption studies and
indications of extended gas around present-day galaxies.   

\item {5.} Baryonic census clearly favors low values of the Hubble constant, essentially
irrespectively of flattening. 

\noindent Further investigations, especially of the epoch of MACHO formation, probed by early
damped Ly$\alpha$ and similar gas-rich systems will be necessary to completely
clear the picture of ramification of the baryonic matter in the universe into
diffuse and collapsed components. This, coupled with microlensing advances, should be
able to finally solve the problem of baryonic component of dark matter and unify
several branches of astrophysical research into a coherent picture of the evolution
of matter, as we know it, in the universe. 

\section*{Acknowledgements}

The authors are happy to express their gratitude to Dr. Geza Gyuk for stimulating comments and
encouragement. Useful discussions with Dr. Slobodan Ninkovi\'c 
and helpful remarks by Dr. Giuliano Giuricin
are also 
acknowledged.
SS acknowledges the financial support of the University of Trieste 
where part of this work was carried out.

}
\bye